\documentclass[aps, preprint, amsmath,amssymb, showpacs,superscriptaddress]{revtex4-1}

\usepackage{graphicx}  
\usepackage{dcolumn}   
\usepackage{bm}        
\usepackage{amssymb}   
\usepackage[english]{babel}
\usepackage[utf8x]{inputenc}
\usepackage{subfigure}
\usepackage{textcomp}
\usepackage{eso-pic}
\usepackage{wasysym}
\usepackage{fancyhdr}
\usepackage{lmodern}
\usepackage{siunitx}
\usepackage{enumitem}
\usepackage{upgreek}
\begin{document}
\title{Femtosecond X-Ray Scattering Study of Ultrafast Photoinduced Structural Dynamics in Solvated [Co(terpy)$_{2}$]$^{2+}$}

\author{Elisa Biasin}
\affiliation{Department of Physics, Technical University of Denmark, Fysikvej 307, DK-2800 Kongens Lyngby, Denmark.}

\author{Tim Brandt van Driel}
\affiliation{Department of Physics, Technical University of Denmark, Fysikvej 307, DK-2800 Kongens Lyngby, Denmark.}
\author{Kasper S. Kj\ae r}
\affiliation{Department of Physics, Technical University of Denmark, Fysikvej 307, DK-2800 Kongens Lyngby, Denmark.}
\affiliation{Department of Chemical Physics, Lund University, Box 118, S-22100 Lund, Sweden.}
\affiliation{PULSE Institute, SLAC National Accelerator Laboratory, Menlo Park, California 94025, USA.}
\author{Asmus O. Dohn}
\affiliation{Department of Chemistry, Technical University of Denmark, Kemitorvet 207, DK-2800 Kongens Lyngby, Denmark.}
\author{Morten Christensen}
\affiliation{Department of Physics, Technical University of Denmark, Fysikvej 307, DK-2800 Kongens Lyngby, Denmark.}

\author{Tobias Harlang}
\affiliation{Department of Chemical Physics, Lund University, Box 118, S-22100 Lund, Sweden.}
\author{Pavel Chabera}
\affiliation{Department of Chemical Physics, Lund University, Box 118, S-22100 Lund, Sweden.}
\author{Yizhu Liu}
\affiliation{Department of Chemical Physics, Lund University, Box 118, S-22100 Lund, Sweden.}
\affiliation{Centre for Analysis and Synthesis, Department of Chemistry, Lund
University, Box 124, Lund SE-22100, Sweden.}
\author{Jens Uhlig}
\affiliation{Department of Chemical Physics, Lund University, Box 118, S-22100 Lund, Sweden.}

\author{M{\'a}ty{\'a}s P{\'a}pai}
\affiliation{Department of Chemistry, Technical University of Denmark, Kemitorvet 207, DK-2800 Kongens Lyngby, Denmark.}
\affiliation{Wigner Research Centre for Physics, Hungarian Academy Sciences, H-1525 Budapest, Hungary.}
\author{Zolt{\'a}n N{\'e}meth}
\affiliation{Wigner Research Centre for Physics, Hungarian Academy Sciences, H-1525 Budapest, Hungary.}

\author{Robert Hartsock}
\affiliation{PULSE Institute, SLAC National Accelerator Laboratory, Menlo Park, California 94025, USA.}
\author{Winnie Liang}
\affiliation{PULSE Institute, SLAC National Accelerator Laboratory, Menlo Park, California 94025, USA.}

\author{Jianxin Zhang}
\affiliation{School of Environmental and Chemical Engineering, Tianjin Polytechnic University, Tianjin 300387, China}

\author{Roberto Alonso-Mori}
\affiliation{LCLS, SLAC National Accelerator Laboratory, Menlo Park, California 94025, USA.}
\author{Matthieu Chollet}
\affiliation{LCLS, SLAC National Accelerator Laboratory, Menlo Park, California 94025, USA.}
\author{James M. Glownia}
\affiliation{LCLS, SLAC National Accelerator Laboratory, Menlo Park, California 94025, USA.}
\author{Silke Nelson}
\affiliation{LCLS, SLAC National Accelerator Laboratory, Menlo Park, California 94025, USA.}
\author{Dimosthenis Sokaras}
\affiliation{LCLS, SLAC National Accelerator Laboratory, Menlo Park, California 94025, USA.}

\author{Tadesse A. Assefa}
\affiliation{European XFEL GmbH, Albert-Einstein-Ring 19, D-22761 Hamburg, Germany.}
\author{Alexander Britz}
\affiliation{European XFEL GmbH, Albert-Einstein-Ring 19, D-22761 Hamburg, Germany.}
\author{Andreas Galler}
\affiliation{European XFEL GmbH, Albert-Einstein-Ring 19, D-22761 Hamburg, Germany.}
\author{Wojciech Gawelda}
\affiliation{European XFEL GmbH, Albert-Einstein-Ring 19, D-22761 Hamburg, Germany.}
\affiliation{Institute of Physics, Jan Kochanowski University, 25-406 Kielce, Poland.}

\author{Christian Bressler}
\affiliation{European XFEL GmbH, Albert-Einstein-Ring 19, D-22761 Hamburg, Germany.}
\author{Kelly J. Gaffney}
\affiliation{PULSE Institute, SLAC National Accelerator Laboratory, Menlo Park, California 94025, USA.}
\author{Henrik T. Lemke}
\affiliation{LCLS, SLAC National Accelerator Laboratory, Menlo Park, California 94025, USA.}
\affiliation{SwissFEL, Paul Scherrer Institut, 5232 Villigen PSI, Switzerland.}
\author{Klaus B. M\o ller}
\affiliation{Department of Chemistry, Technical University of Denmark, Kemitorvet 207, DK-2800 Kongens Lyngby, Denmark.}
\author{Martin M. Nielsen}
\affiliation{Department of Physics, Technical University of Denmark, Fysikvej 307, DK-2800 Kongens Lyngby, Denmark.}
\author{Villy Sundstr{\"o}m}
\affiliation{Department of Chemical Physics, Lund University, Box 118, S-22100 Lund, Sweden.}
\author{Gy{\"o}rgy Vank{\'o}}
\affiliation{Wigner Research Centre for Physics, Hungarian Academy Sciences, H-1525 Budapest, Hungary.}
\author{Kenneth W{\"a}rnmark}
\affiliation{Centre for Analysis and Synthesis, Department of Chemistry, Lund
University, Box 124, Lund SE-22100, Sweden.}
\author{Sophie E. Canton}
\affiliation{IFG Structural Dynamics of (Bio)chemical Systems, Max Planck Institute for Biophysical Chemistry, Am Fassberg 11, D-37077 Goettingen, Germany}
\affiliation{FS-SCS, Structural Dynamics with Ultra-short Pulsed X-rays, Deutsches Elektronen-Synchrotron (DESY), Notkestrasse 85, D-22607 Hamburg, Germany}
\author{Kristoffer Haldrup}
\affiliation{Department of Physics, Technical University of Denmark, Fysikvej 307, DK-2800 Kongens Lyngby, Denmark.}

\begin{abstract}
We study the structural dynamics of photoexcited [Co(terpy)$_2$]$^{2+}$ in an aqueous solution with ultrafast x-ray diffuse scattering experiments conducted at the Linac Coherent Light Source. Through direct comparisons with density functional theory calculations, our analysis shows that the photoexcitation event leads to elongation of the Co-N bonds, followed by coherent Co-N bond length oscillations arising from the impulsive excitation of a vibrational mode dominated by the symmetrical stretch of all six Co-N bonds. This mode has a period of 0.33 ps and decays on a subpicosecond time scale. We find that the equilibrium bond-elongated structure of the high spin state is established on a single-picosecond time scale and that this state has a lifetime of $\sim$ 7 ps.
\end{abstract}

\maketitle

Several Co(II) compounds are known to transition between their low spin (LS) and high spin (HS) electronic states~\cite{goodwin2004spin, krivokapic2007, enachescu2007optical}. Such transitions can be induced by temperature increase, excitation by light or high magnetic fields~\cite{bousseksou2002}, and they are accompanied by distinct changes in magnetic and structural properties that may be exploited in the design of display and memory devices~\cite{application1, application2} and in single-molecule spintronic applications~\cite{eliseoruiz2014}. The realization of exploitable spin-state transitions (SSTs) in Co(II) compounds is more challenging than in the corresponding Fe(II) complexes, which have been investigated in great detail during the last decades~\cite{gutlich2004spin, chergui, cammarata2014prl, vanko2006jpcb, canton2014ani, Nance2015, Haldrup2016_LCLS2011,  lemkeAr}. These challenges stem from the partial occupation of the antibonding $e_g^*$ orbitals in the ground state, which leads to smaller structural changes arising from the SST phenomenon; the corresponding smaller energy barriers between the potential surfaces of the HS and LS Co(II) states result in faster dynamics~\cite{goodwin2004spin}, as well as a high sensitivity to the crystalline environment or to the solvent properties~\cite{krivokapic2007}. The key structural parameters for the SSTs are the Co-N bond lengths~\cite{gutlich2004spin}, but the time scales and the dynamics of the LS-HS transitions have remained unclear for Co-compounds. Time-resolved x-ray scattering can be used to monitor such structural changes and dynamics if the time resolution of the experiment is sufficiently high. X-ray free electron lasers (XFELs) provide ultra-short ($\sim$ 30 fs) x-ray pulses and high flux allowing the nuclear dynamics following photoexcitation to be recorded at the required femtosecond time scales~\cite{lemke, xpp}. Here we report, for the first time, direct measurements of the excited-state structure and the ultrafast structural dynamics of a solvated Co(II) complex upon a photoinduced SST.

\begin{figure}
\includegraphics[scale=0.15]{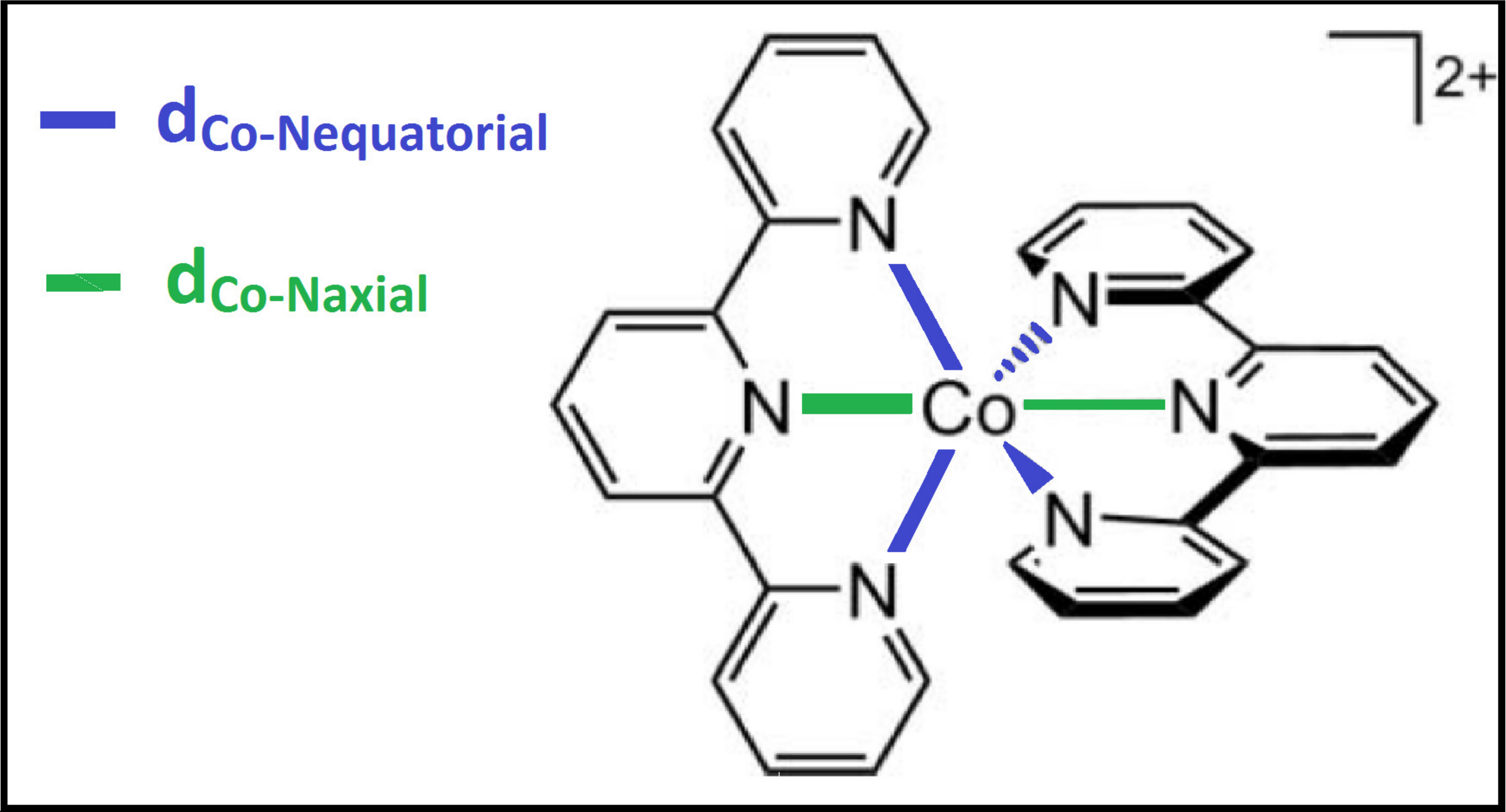}
\caption{\label{spectra} Schematic representation of the structure of [Co(terpy)$_2$]$^{2+}$. The LS → HS transition can be induced by photoexcitation with a 530 nm laser pulse and is characterized by an anisotropic expansion of the metal-ligand bonds. Axial and equatorial Co-N bonds are highlighted in different colors.}
\end{figure}

Fig.~\ref{spectra} shows the molecular structure of [Co(terpy)$_{2}$]$^{2+}$ (terpy = 2,2´:6´, 2´´-terpyridine). In this six-coordinated complex, the $d^{7}$ Co center can be either a LS doublet state or a HS quartet state~\cite{krivokapic2007, Kremer}. In solid-state samples, the relative populations of both spin states depend strongly on the temperature and on the crystalline environment~\cite{JudgeBaker1967, hayami2011SCOterpy, oshio}. In crystallographic studies the compound was observed to be compressed in the LS state (short axial and long equatorial Co-N bonds) due mostly to the geometrical constraints of the coordinating tridentate ligands, and may also exhibit asymmetry, with one ligand being closer to the Co center than the other due to a pseudo Jahn-Teller effect~\cite{hayami2011SCOterpy, Vargas}. Upon LS → HS transition in solid-state samples, the axial bond length has been observed to increase by up to 0.21 Å and the equatorial by 0.07 Å, depending on the anion and the degree of hybridization~\cite{crystalStructure}. As reported by Vargas $et$ $ al$.~\cite{Vargas}, density functional theory (DFT) calculations in the gas phase also predict an anisotropic increase of the Co-N bonds upon the LS → HS spin change (an increase of 0.16 Å and 0.05 Å for the axial and equatorial bonds, respectively). A few studies on the properties of [Co(terpy)$_2$]$^{2+}$ in solution also exist~\cite{enachescu2007optical, beattie1995dynamics, Kremer, krivokapic2007}. Kremer $et$ $al$.~\cite{Kremer} report that solvated [Co(terpy)$_2$]$^{2+}$  is predominantly LS at room temperature, and Enachescu $et$ $al$. demonstrated that photoexcitation in the visible range populates the metal to ligand charge transfer (MLCT) state from which the HS state is populated~\cite{enachescu2007optical}. Very little information is available regarding the excited-state decay pathways and the HS → LS relaxation time is currently only known to be less than 2 ns~\cite{beattie1995dynamics}. 

In this work, we utilized x-ray diffuse scattering (XDS) laser pump/x-ray probe experiments to study the formation, structure and decay of the HS state of aqueous [Co(terpy)$_{2}$]$^{2+}$. The measurements were conducted at the x-ray pump-probe (XPP) instrument at the Linac Coherent Light Source (LCLS) XFEL facility~\cite{xpp}. A 20 mM aqueous solution of [Co(terpy)$_{2}$]$^{2+}$ was pumped trough a nozzle producing a 100 $\upmu$m liquid sheet flowing in the vertical direction at a flow rate sufficient to fully replace the sample between successive pump/probe events. The photocycle was initiated by 70 $\upmu$J laser pulses at 530 nm and with a 70 fs pulse width (FWHM), focused onto a spot of 150 $\upmu$m (FWHM). The 8.3 keV x-ray probe pulses overlapped with the pump laser at the sample position. The time delay t between the laser and the x-ray pulses was determined for every pump-probe event with $\sim$ 10 fs (FWHM) resolution using the XPP timing tool~\cite{timingtoolXPP}. The scattered x-rays were detected by a Cornell-SLAC pixel array detector~\cite{cspad} 70 mm after the sample, covering scattering vectors Q up to 3.5 Å$^{-1}$. 

\begin{figure*}
\includegraphics[scale = 0.22]{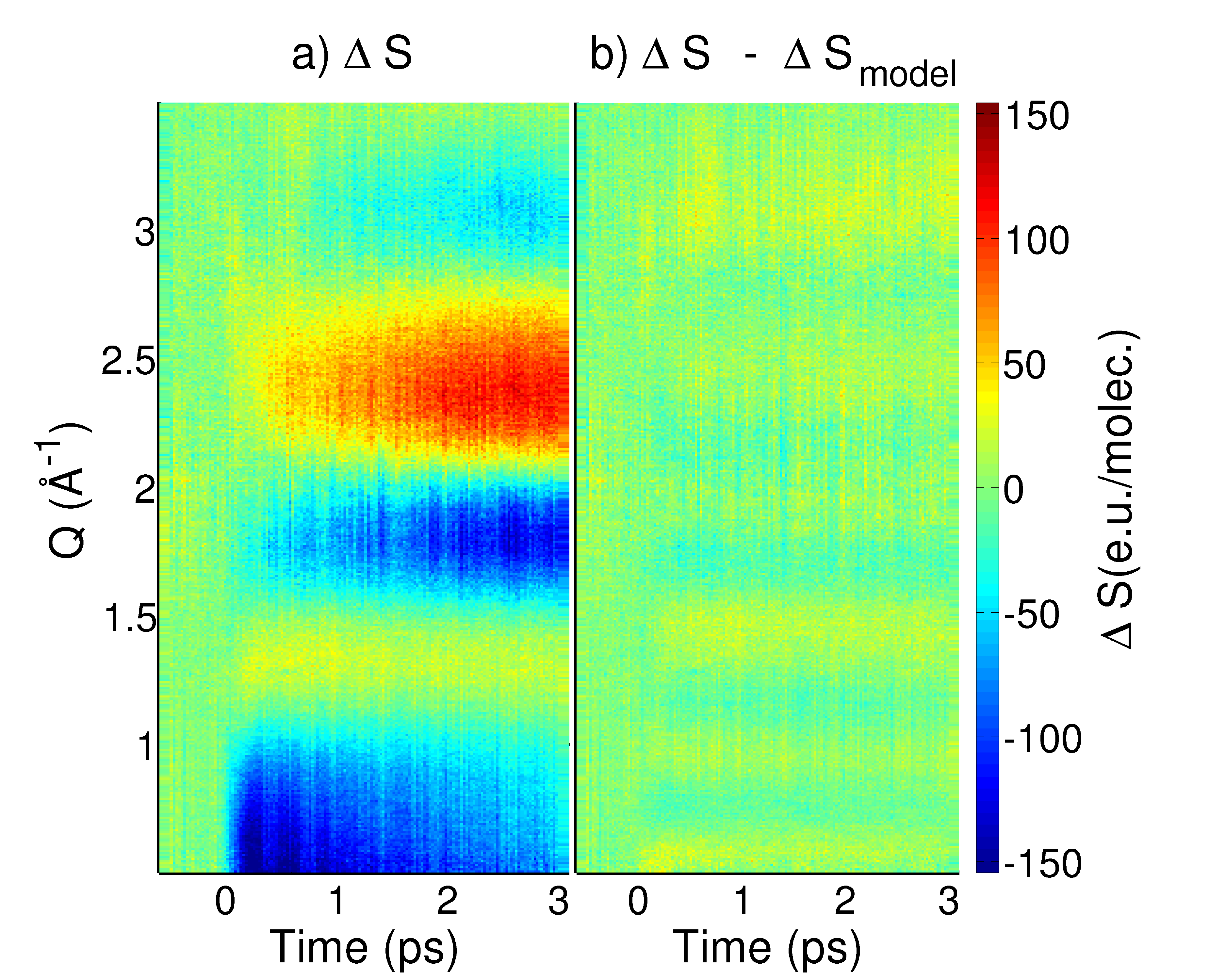} 
\includegraphics[scale = 0.07]{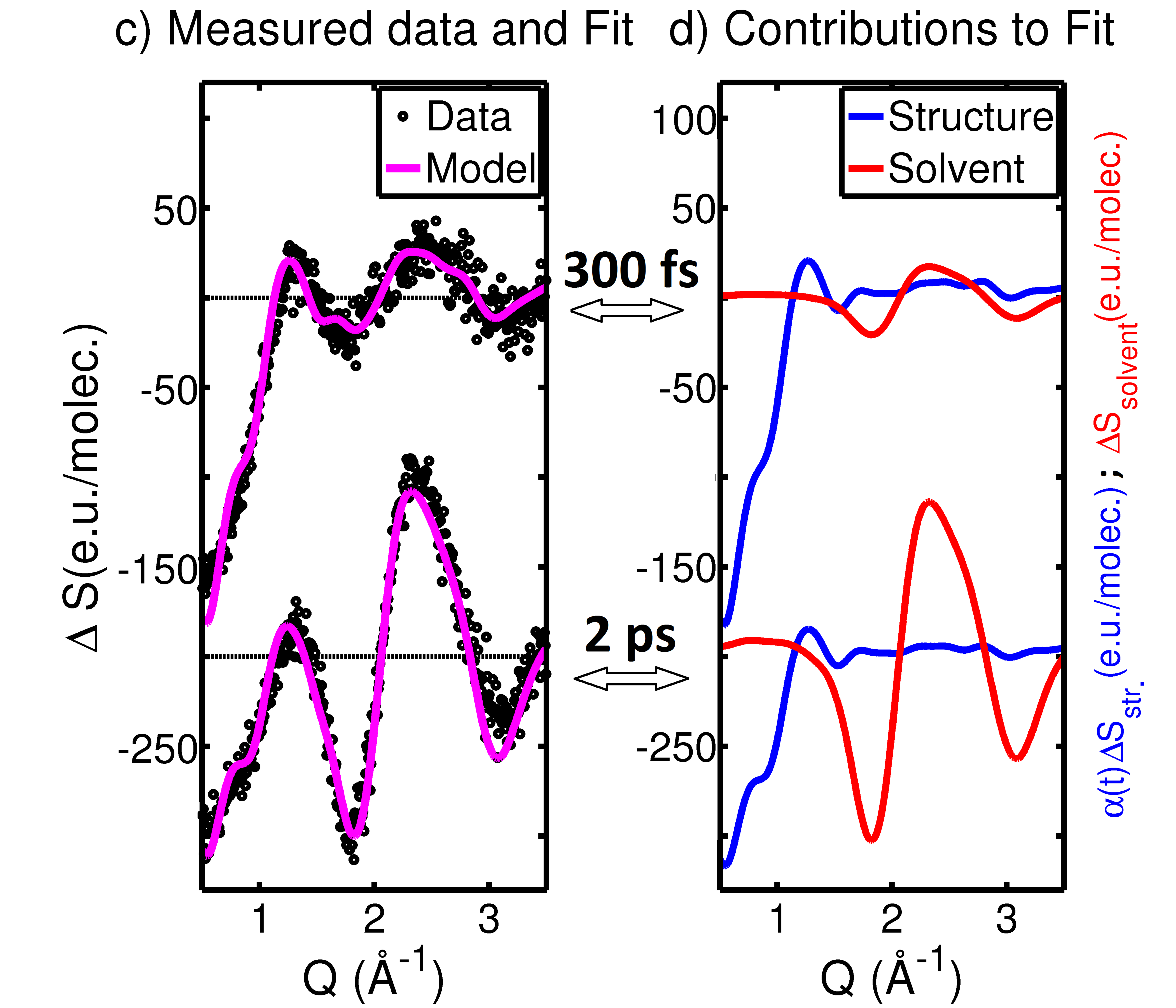}
\caption{\label{fit} \textbf{(a)} Measured difference scattering signal ($\Delta$S) of photoexcited [Co(terpy)$_2$]$^{2+}$ in water. \textbf{(b)} Residuals obtained by subtracting the model ($\Delta$S$_{\text{model}}$) from the experimental data. \textbf{(c),(d)} Fit of the 1D difference scattering curves at 300 fs and 2 ps. (c) The modelled difference signal (the magenta line) overlaid over the experimental data (black points). (d) The contributions to the model from the structural changes (solute and cage, the blue line) and from the bulk solvent (the red line).}
\end{figure*}

Following detector corrections~\cite{timSVD}, the scattering signal was scaled to the liquid unit cell reflecting the stoichiometry of the sample~\cite{haldrup2010analysis}, yielding the acquired signal in electron units per solute molecule (e.u./molec.). Individual 2D difference scattering patterns were obtained by subtracting images where the pump laser was dropped before the sample from those where the pump laser had interacted with the sample. The patterns were then time sorted and averaged in $\sim$ 23 fs wide bins. Finally, 1D isotropic and anisotropic difference scattering signals were extracted~\cite{aniUlf}. Fig.~\ref{fit}(a) shows the measured isotropic difference signals $\Delta$S(Q, t) in a 2D representation.

$\Delta$S(Q,t) can be considered as arising from three contributions~\cite{Kreuff-Fe}: the structural changes in the solute molecules ($\Delta$S$_{\text{solute}}$), the local changes in geometry and rearrangements of the solvent molecules in close proximity to the solute ($\Delta$S$_{\text{cage}}$), and the temperature and density changes in the bulk solvent following energy deposition ($\Delta$S$_{\text{solvent}}$). 

$\Delta$S$_{\text{solute}}$(Q) can be directly calculated from putative structural models of the molecule through the Debye equation (Eq.~S2 in the Supplemental Material (SM)~\cite{sm}). As a starting point for the present analysis the difference scattering signal expected upon the photoexcitation was calculated from the LS and HS DFT-optimized geometries of [Co(terpy)$_2$]$^{2+}$:
\begin{equation}
\Delta \text{S}_{\text{solute}}(\text{Q}) = \text{S}_{\text{HS}}(\text{Q}) - \text{S}_{\text{LS}}(\text{Q}),
\label{deltaS} 
\end{equation}
The DFT calculations were carried out as detailed in the SM~\cite{sm} and Table~\ref{tab1} reports the key DFT-calculated structural parameters. Upon the LS → HS transition the Co-N bonds expand $\sim$ 0.16 Å and $\sim$ 0.08 Å along the axial and the equatorial directions, respectively, in good agreement with the earlier study by Vargas $et$ $al$.~\cite{Vargas}. The ratio between the axial and the equatorial Co-N distances is defined as $\eta$. In the LS state the average $\eta$ is 0.91 (0.88 and 0.92 for the two ligands, with the difference due to the Jahn-Teller effect), while in the HS state $\eta$ increases to 0.95 (for both ligands).  

\begin{table}
\caption{\label{tab1} Structural parameters of the DFT-calculated LS and HS structures of [Co(terpy)$_2$]$^{2+}$ obtained in the present study. d$_{\text{Co-N,axial}}$ and d$_{\text{Co-N,equatorial}}$  are averages over the two axial  and the four equatorial metal-ligand bond distances, respectively, and $\eta = \frac{\text{d}_{\text{Co-Naxial}}}{\text{d}_{\text{Co-Nequatorial}}}$. The change of each parameter upon the LS → HS spin transition is also reported and compared with the values obtained from the measured data.}
\begin{ruledtabular}
\begin{tabular}{ccc|ccc}
& LS & HS & & DFT & Measured \\\hline
d$_{\text{Co-Naxial}}$ (Å) & 1.902 & 2.058 & $\Delta$d$_{\text{Co-Naxial}}$ (Å) & 0.16 & 0.13 \\
d$_{\text{Co-Nequatorial}}$ (Å) & 2.08 & 2.16 & $\Delta$d$_{\text{Co-Nequatorial}}$ (Å) & 0.08 & 0.06 \\\hline
$\eta$ & 0.91 & 0.95 &  &  & \\
\end{tabular}
\end{ruledtabular}
\end{table}

The cage contribution $\Delta$S$_{\text{cage}}$(Q) to the simulated signal was calculated from the Radial Distribution Functions of the solute-solvent atom pairs~\cite{asmusRDF} determined through classical molecular dynamics (MD) simulations~\cite{sm}. The contribution from changes in the solute structure and solvation cage are related 1:1 and can therefore be combined under the term ``structure'', $\Delta$S$_{\text{str.}}$(Q), such that:
\begin{equation}
\Delta \text{S}_{\text{str.}}(\text{Q}) = \Delta \text{S}_{\text{solute}}(\text{Q}) + \Delta \text{S}_{\text{cage}}(\text{Q}).
\label{structure}
\end{equation}

Finally, the bulk-solvent term $\Delta$S$_{\text{solvent}}$(Q) has been shown to be very well described by a linear combination of solvent difference signals, $\frac{\partial \text{S(Q)}}{\partial \text{T} }\bigg|_{\rho}$ and $\frac{\partial \text{S(Q)}}{\partial \rho} \bigg|_{\text{T}}$, which can be measured in separate experiments~\cite{cammarata2006solvent, KasperSolvent}:
\begin{equation}
\Delta \text{S(Q,t)}_{\text{solvent}} =  \Delta \text{T(t)} \frac{\partial \text{S(Q)}}{\partial \text{T} }\bigg|_{\rho} + \Delta\rho(\text{t}) \frac{\partial \text{S(Q)}}{\partial \rho} \bigg|_{\text{T}},
\label{solvent}
\end{equation}
where $\Delta$T and $\Delta \rho$ are the changes in temperature and density, respectively. Such solvent differentials for XDS experiments are archived for a range of solvents~\cite{KasperSolvent, URLsolvent} and are used in the present work. In contrast to earlier experiments on Fe SST compounds~\cite{Kreuff-Fe} we observe no density change above our detection limit of 0.05 kg/m$^3$ (Fig.~S1(b)~\cite{sm}) and this term was thus excluded from the analysis. 

From visual inspection of the measured difference signal in Fig.~\ref{fit}(a), we qualitatively observe a very fast rise of a negative feature at low-Q (Q$<$1 Å$^{-1}$) which gradually decays over the course of several picoseconds. Such a low-Q feature is characteristic of an increase in the solute size. On the few picosecond time scale, a distinct signal around Q = 2 Å$^{-1}$ grows in. This feature is identified as the characteristic difference signal arising from a temperature increase of the aqueous solvent. In the low-Q region, oscillatory features as a function of time can be observed and indicate structural dynamics along the main coordinate of the structural changes, in the present case the Co-N bond lengths (d$_{\text{Co-N}}$). The latter is therefore introduced as a time-dependent parameter in Eq.~\ref{deltaS}: 
\begin{equation}
\text{S}_{\text{HS}}\text{(Q, t)} = \text{S}_{\text{HS}}(\text{Q}, \text{d}_{\text{Co-N}}(\text{t})).
\label{sHS}
\end{equation}
Specifically, d$_{\text{Co-Naxial}}$ of the HS structure was allowed to vary $\pm$ 0.1 Å from the value reported in Table~\ref{tab1} while the ratio $\eta$, through which d$_{\text{Co-Nequatorial}}$ can be calculated and included in the structural modeling, was kept fixed to 0.95 in the analysis. Thus all six Co-N bond length changes are parametrized through the single structural parameter d$_{\text{Co-Naxial}}$.

Based on the considerations outlined above, the full model applied to fit and interpret the measured difference signal is thus:
\begin{equation}
\Delta \text{S}_{\text{model}} \text{(Q,t)} = \alpha(\text{t})\Delta \text{S}_{\text{str.}}\text{(Q, t)} + \Delta \text{T} (\text{t}) \frac{\partial \text{S} (\text{Q})}{\partial \text{T} } \bigg|_{\rho}
\label{model}
\end{equation} where $\alpha$(t) describes the time-dependent excitation fraction of the solute, which in the context of the present analysis is assumed to be described by an exponential decay starting at t$_0$, i.e. the arrival time of the laser pump. The time resolution of the experiment is included by convolution with the (Gaussian) instrument response function (IRF) to yield the following expression for $\alpha$(t):
\begin{equation}
\alpha(\text{t}) = \text{IRF}(\sigma_{\text{IRF}}, \text{t} ) \otimes H(\text{t}-\text{t}_0) A e^{-\frac{\text{t} - \text{t}_0}{\tau}}
\label{convy}
\end{equation}
where $\sigma_{\text{IRF}}$ is the width of the IRF, $A$ and $\tau$ are the amplitude and the lifetime of the exponential function representing, respectively, the initial excitation fraction and the lifetime of the bond-elongated excited-state, and $H$ is the Heaviside step function centered at t$_0$ (as detailed in Eq.~S3~\cite{sm}). We note that assuming the excited-state population to be given by the integral of a Gaussian envelope of the excitation pulse is an approximation, especially given the high intensity of the optical excitation, as discussed in further detail below. $\sigma_{\text{IRF}}$ and t$_0$ were determined from the transient solvent contribution to the anisotropic part of the difference scattering signal (Fig.~S4~\cite{sm}) from which we find $\sigma_{\text{IRF}}$ = 0.05 ps $\pm$ 0.03 ps. Furthermore, we  estimated the lifetime of the HS state from a single set of measurements out to 20 ps. The analysis of this data set is presented in the SM and yields $\tau$ = 6.8 ps $\pm$ 0.8 ps ( Fig.~S8(a)~\cite{sm}), allowing us to constrain this parameter in Eq.~\ref{convy}.
 
From these considerations, the number of free parameters in the model described by Eq.~\ref{model} are reduced to three: $A$, d$_{\text{Co-Naxial}}$ and $\Delta$T. The model was fitted to the acquired difference signal $\Delta$S(Q) for all time delays simultaneously within a standard  $\chi^{2}$  (Eq.~S6~\cite{sm}) minimization framework~\cite{Jun_Ihee_2010}. Good fits were observed for all time delays and Fig.~\ref{fit}(b) shows the residuals after subtracting the model from the measured data. Fig.~\ref{fit}(c and d) shows examples of the fitting results at two time delays, 300 fs and 2 ps. 
\begin{figure}
\includegraphics[scale = 0.120]{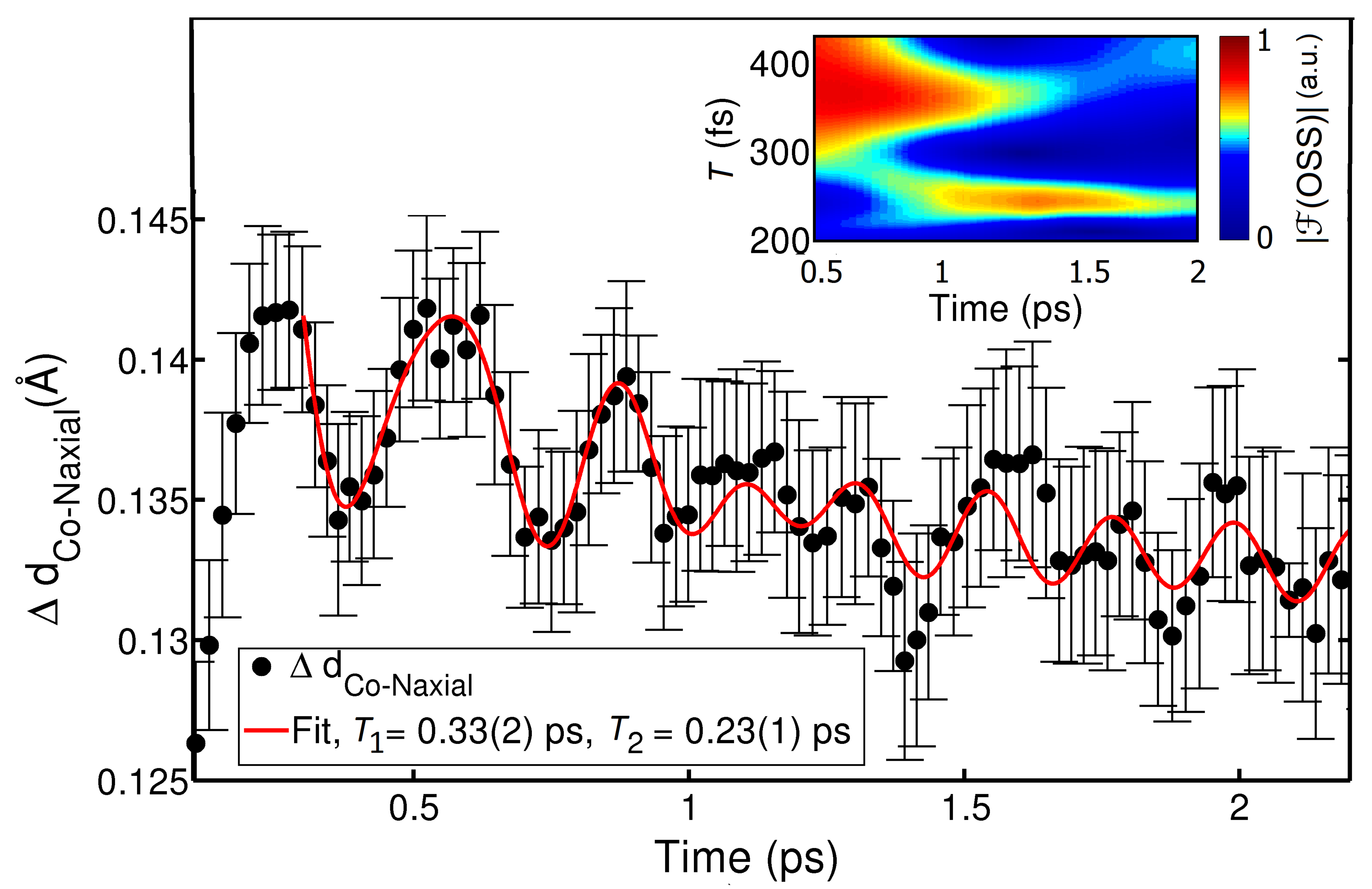}
\caption{\label{bonds} Time evolution of the Co-N bond lengths (black dots) upon photoexcitation, smoothed with 4-point ($\sim$  100 fs) moving average filter. The insert shows a time-resolved Fourier transform of the oscillatory part of the difference  scattering signal (Fig.~\ref{fit}(a) and Fig.~S6~\cite{sm}), indicating sequential activation of two vibrational modes. The red line shows a heuristic fit, incorporating sequential activation of first a $T_{1}\sim$ 0.33 ps mode and then a  $T_{2}\sim$ 0.23 ps mode identified as, respectively, breathing- and pincerlike by direct comparison with our DFT calculations.}
\end{figure}

From the kinetics part of the fit of our model to the acquired data, the initial excitation fraction $A$ was found to be 34 \% $\pm$ 2 \%. Regarding the difference signal arising from solvent heating, the analysis of $\Delta$T(t) is discussed in detail in the SM (Fig.~S3 and S8(b)~\cite{sm}) but briefly, it is found to be well described by a  broadened double exponential dominated ($>$ 90\%) by a response with a grow-in time constant of 4.0 ps $\pm$ 0.6 ps. A total solvent temperature increase $\Delta$T = 0.8 K is found, which is 0.4 K more than the amount of energy expected to be released through non-radiative decay processes after single-photon excitation of the solute. As detailed in the SM~\cite{sm}, this extra heat can be ascribed to multi-photon absorption due to the relatively high excitation laser intensity and short pulse length. A direct comparison with data taken at 3 times lower laser power ( Fig.~S10~\cite{sm}) shows that the multi-photon absorption has no discernible impact on the structural response of the solute molecules. 

Turning to the key results of this Letter, Fig.~\ref{bonds} shows the best-fit result for the changes in d$_{\text{Co-Naxial}}$ from the ground to the excited-state as a function of time (the black data points). Following excitation, the axial Co-N bond increases by $\Delta$d$_{\text{Co-Naxial}}$ = 0.14 Å and exhibits oscillations. On the 1 ps time scale, the axial Co-N bond length of the excited-state ensemble decreases by $\sim$ 0.01 Å and then remains constant over the $\sim$ 7 ps lifetime of the HS state. Thus, d$_{\text{Co-Naxial}}$ and d$_{\text{Co-Nequatorial}}$ are found to be, respectively, 0.13 Å and 0.06 Å longer in the HS state than in the LS state; distance changes which are slightly smaller than the DFT predictions (Tab.\ref{tab1}). The rise time of the solvent heating signal indicates that full thermal equilibration with the surrounding solvent takes place in about 4 ps.
 
The inset of Fig.~\ref{bonds} shows a time-dependent Fourier transform ($\mathcal{F}$) of the oscillatory structural signal (OSS) contained in $\Delta S$ and calculated as detailed in the SM~\cite{sm}. From this, we observe two distinct components: one mode which appears within the time resolution of our experiment and decays on a $\sim$ 1 ps time scale, and one mode which grows in after 1 ps. The red line in Fig.~\ref{bonds} illustrates the fit of a heuristic model to the data after the initial lengthening of the Co-N axial bond. The model is comprised of two sinusoidals ( Eq.~S4~\cite{sm}), the first one being damped and driving the second. Both sinusoidals are broadened by the IRF and superimposed on an exponentially decaying background with a time constant of 0.7 ps $\pm$ 0.1 ps. From this fit, we find that the period $T_{1}$ of the main oscillation is 0.33 ps $\pm$ 0.03 ps and that the damping time is 0.4 ps $\pm$ 0.1 ps. On the same time scale we observe the growing of the second oscillation with a period of $T_{2}$ = 0.23 ps $\pm$ 0.01 ps. In the framework of this analysis and by direct comparison with the DFT-calculated vibrational modes of the HS state, we assign the first component to a breathinglike mode (Movie S1~\cite{sm}) with synchronous stretching of all six Co-N bonds, whereas the second, weaker component is assigned to arise from a pincerlike movement of the tridentate ligands (Movie S2~\cite{sm}). The assignment of these modes is in good agreement with the recent work on related Fe(II) complexes~\cite{cammarata2014prl, Majed-FeBpy2015, consani2009, lemkeAr}, where the immediately excited stretching modes were quickly damped as energy was dissipated into other degrees of freedom. Future experiments utilizing higher x-ray energies to access a larger region of momentum space should facilitate detailed studies of the structural degrees of freedom (as recently demonstrated for [Fe(terpy)$_2$]$^{2+}$ on synchrotron time scales (100 ps)~\cite{vanko, canton2014ani, JTcanton2015}) involved in the structural relaxation of the electronically excited-state. Such studies may be fruitfully combined with ab initio MD~\cite{dohn2014}, thus going beyond the classical-mechanics description of the combined DFT/MD simulations used in the present analysis.

Returning to the solute dynamics, by assuming that the excited-state potential is well approximated by a harmonic potential and if the population of this state is nearly instantaneous, one would expect the ensemble mean of the Co-N bond length to reach its maximum value one half period ($\sim$ 0.17 ps) after excitation. From Fig.~\ref{bonds}, we find this point to be reached only after 0.25 ps. By singular value decomposition of the structural contribution to $\Delta$S (Figs.~S4 and S5~\cite{sm}), we find this observation of a delayed structural transition to be model independent and further find that the delayed onset is well described by an exponential grow-in ($\tau$ = 0.06 ps $\pm$ 0.01 ps) of the signal with a 0.08 ps $\pm$ 0.02 ps phase shift of the oscillations. These observations are consistent with the excited-state structural dynamics taking place on several potential surfaces: photoexcitation produces a MLCT excited-state, while bond elongation is believed to occur predominantly in the metal-centered HS excited-state. Referring back to the discussion of Eq.~\ref{convy}, we note that this expression is only strictly applicable in a regime of linear response and that, therefore, the $\sim$ 100 fs delay in bond elongation can be considered only a coarse, structural, measure of the time scale involved in the electronic processes of intersystem crossing and internal conversion that eventually leads to formation of the HS state. This delay, while sufficiently fast to launch the observed synchronous Co-N stretch mode, leads to significant broadening of the HS population in terms of the Co-N bond lengths. This in turn leads to the observed phase shift and the comparatively low amplitude of the observed oscillations.

These results demonstrate how time-resolved x-ray scattering with solution-state samples can be utilized to accurately characterize femtosecond structural dynamics as photoexcited molecules traverse the potential energy landscape of the excited-state(s). We believe the results and methodology presented here to be broadly applicable, and we envision that these types of experiments will have a significant impact on our understanding of the fundamental mechanisms at work in SST systems and in both natural and artificial photosensitizers, where the redistribution of energy to different and strongly coupled internal degrees of freedom (both electronic and structural) are of key importance. 

\section*{Acknowledgements}
The DTU-affiliated authors would like to gratefully acknowledge DANSCATT for funding the beam time efforts. M.M.N., K.B.M., K.H., A.D., E.B., K.S.K. and T.B.V.D. acknowledges support from Te Frie Forskningsråd (DFF) and the Lundbeck Foundation. S.E.C acknowledges funding from SFB 1073. This project was also supported by the European Research Council via Contract
ERC-StG-259709, the Hungarian Scientific Research Fund (OTKA) under Contract K29724 and the Lend{\"u}let (Momentum) Program of the Hungarian
Academy of Sciences (LP2013-59). Z.N. acknowledges support from the Bolyai
Fellowship of the Hungarian Academy of Sciences. M.P. acknowledges support from the People Programme (Marie Curie Actions) of the European Union’s Seventh Framework Programme (FP7/2007-2013) under REA Grant Agreement 609405 (COFUNDPostdocDTU). J.U. gratefully acknowledges the continued funding from the Knut and
Alice Wallenberg Foundation. J.Z. greatly acknowledges support from NSFC (21302138) and Tianjin High School Science and Technology Fund Planning Project (20130504). R.W.H., H.W.L. and K.J.G. acknowledge support from the AMOS program within the Chemical Science, Geosciences and Biosciences Division of the Office of Basic Energy Sciences, Office of Science, U.S. Department of Energy. T.A., A.B., W.G., A.G., and C.B. acknowledge funding from the German Science Foundation (DFG) via SFB925 and from the Centre of Ultrafast Imaging (CUI). Use of the Linac Coherent Light Source (LCLS), SLAC National Accelerator Laboratory, is supported by the U.S. Department of Energy, Office of Science, Office of Basic Energy Sciences under Contract No. DE-AC02-76SF00515.

\bibliography{references}{}

\begin{thebibliography}{41}%
\makeatletter
\providecommand \@ifxundefined [1]{%
 \@ifx{#1\undefined}
}%
\providecommand \@ifnum [1]{%
 \ifnum #1\expandafter \@firstoftwo
 \else \expandafter \@secondoftwo
 \fi
}%
\providecommand \@ifx [1]{%
 \ifx #1\expandafter \@firstoftwo
 \else \expandafter \@secondoftwo
 \fi
}%
\providecommand \natexlab [1]{#1}%
\providecommand \enquote  [1]{``#1''}%
\providecommand \bibnamefont  [1]{#1}%
\providecommand \bibfnamefont [1]{#1}%
\providecommand \citenamefont [1]{#1}%
\providecommand \href@noop [0]{\@secondoftwo}%
\providecommand \href [0]{\begingroup \@sanitize@url \@href}%
\providecommand \@href[1]{\@@startlink{#1}\@@href}%
\providecommand \@@href[1]{\endgroup#1\@@endlink}%
\providecommand \@sanitize@url [0]{\catcode `\\12\catcode `\$12\catcode
  `\&12\catcode `\#12\catcode `\^12\catcode `\_12\catcode `\%12\relax}%
\providecommand \@@startlink[1]{}%
\providecommand \@@endlink[0]{}%
\providecommand \url  [0]{\begingroup\@sanitize@url \@url }%
\providecommand \@url [1]{\endgroup\@href {#1}{\urlprefix }}%
\providecommand \urlprefix  [0]{URL }%
\providecommand \Eprint [0]{\href }%
\providecommand \doibase [0]{http://dx.doi.org/}%
\providecommand \selectlanguage [0]{\@gobble}%
\providecommand \bibinfo  [0]{\@secondoftwo}%
\providecommand \bibfield  [0]{\@secondoftwo}%
\providecommand \translation [1]{[#1]}%
\providecommand \BibitemOpen [0]{}%
\providecommand \bibitemStop [0]{}%
\providecommand \bibitemNoStop [0]{.\EOS\space}%
\providecommand \EOS [0]{\spacefactor3000\relax}%
\providecommand \BibitemShut  [1]{\csname bibitem#1\endcsname}%
\let\auto@bib@innerbib\@empty
\bibitem [{\citenamefont {Goodwin}(2004)}]{goodwin2004spin}%
  \BibitemOpen
  \bibfield  {author} {\bibinfo {author} {\bibfnamefont {H.~A.}\ \bibnamefont
  {Goodwin}},\ }\href@noop {} {\bibfield  {journal} {\bibinfo  {journal}
  {Topics in Current Chemistry}\ }\textbf {\bibinfo {volume} {234}},\ \bibinfo
  {pages} {23} (\bibinfo {year} {2004})}\BibitemShut {NoStop}%
\bibitem [{\citenamefont {Krivokapic}\ \emph {et~al.}(2007)\citenamefont
  {Krivokapic}, \citenamefont {Zerara}, \citenamefont {Daku}, \citenamefont
  {Vargas}, \citenamefont {Enachescu}, \citenamefont {Ambrus}, \citenamefont
  {Tregenna-Piggott}, \citenamefont {Amstutz}, \citenamefont {Krausz},\ and\
  \citenamefont {Hauser}}]{krivokapic2007}%
  \BibitemOpen
  \bibfield  {author} {\bibinfo {author} {\bibfnamefont {I.}~\bibnamefont
  {Krivokapic}}, \bibinfo {author} {\bibfnamefont {M.}~\bibnamefont {Zerara}},
  \bibinfo {author} {\bibfnamefont {M.~L.}\ \bibnamefont {Daku}}, \bibinfo
  {author} {\bibfnamefont {A.}~\bibnamefont {Vargas}}, \bibinfo {author}
  {\bibfnamefont {C.}~\bibnamefont {Enachescu}}, \bibinfo {author}
  {\bibfnamefont {C.}~\bibnamefont {Ambrus}}, \bibinfo {author} {\bibfnamefont
  {P.}~\bibnamefont {Tregenna-Piggott}}, \bibinfo {author} {\bibfnamefont
  {N.}~\bibnamefont {Amstutz}}, \bibinfo {author} {\bibfnamefont
  {E.}~\bibnamefont {Krausz}}, \ and\ \bibinfo {author} {\bibfnamefont
  {A.}~\bibnamefont {Hauser}},\ }\href {\doibase 10.1016/j.ccr.2006.05.006}
  {\bibfield  {journal} {\bibinfo  {journal} {Coordination Chemistry Reviews}\
  }\textbf {\bibinfo {volume} {251}},\ \bibinfo {pages} {364 } (\bibinfo {year}
  {2007})}\BibitemShut {NoStop}%
\bibitem [{\citenamefont {Enachescu}\ \emph {et~al.}(2007)\citenamefont
  {Enachescu}, \citenamefont {Krivokapic}, \citenamefont {Zerara},
  \citenamefont {Real}, \citenamefont {Amstutz},\ and\ \citenamefont
  {Hauser}}]{enachescu2007optical}%
  \BibitemOpen
  \bibfield  {author} {\bibinfo {author} {\bibfnamefont {C.}~\bibnamefont
  {Enachescu}}, \bibinfo {author} {\bibfnamefont {I.}~\bibnamefont
  {Krivokapic}}, \bibinfo {author} {\bibfnamefont {M.}~\bibnamefont {Zerara}},
  \bibinfo {author} {\bibfnamefont {J.~A.}\ \bibnamefont {Real}}, \bibinfo
  {author} {\bibfnamefont {N.}~\bibnamefont {Amstutz}}, \ and\ \bibinfo
  {author} {\bibfnamefont {A.}~\bibnamefont {Hauser}},\ }\href@noop {}
  {\bibfield  {journal} {\bibinfo  {journal} {Inorganica Chimica Acta}\
  }\textbf {\bibinfo {volume} {360}},\ \bibinfo {pages} {3945} (\bibinfo {year}
  {2007})}\BibitemShut {NoStop}%
\bibitem [{\citenamefont {Bousseksou}\ \emph {et~al.}(2002)\citenamefont
  {Bousseksou}, \citenamefont {Boukheddaden}, \citenamefont {Goiran},
  \citenamefont {Consejo}, \citenamefont {Boillot},\ and\ \citenamefont
  {Tuchagues}}]{bousseksou2002}%
  \BibitemOpen
  \bibfield  {author} {\bibinfo {author} {\bibfnamefont {A.}~\bibnamefont
  {Bousseksou}}, \bibinfo {author} {\bibfnamefont {K.}~\bibnamefont
  {Boukheddaden}}, \bibinfo {author} {\bibfnamefont {M.}~\bibnamefont
  {Goiran}}, \bibinfo {author} {\bibfnamefont {C.}~\bibnamefont {Consejo}},
  \bibinfo {author} {\bibfnamefont {M.-L.}\ \bibnamefont {Boillot}}, \ and\
  \bibinfo {author} {\bibfnamefont {J.-P.}\ \bibnamefont {Tuchagues}},\ }\href
  {\doibase 10.1103/PhysRevB.65.172412} {\bibfield  {journal} {\bibinfo
  {journal} {Physical Review B}\ }\textbf {\bibinfo {volume} {65}},\ \bibinfo
  {eid} {172412} (\bibinfo {year} {2002})}\BibitemShut {NoStop}%
\bibitem [{\citenamefont {Kahn}\ and\ \citenamefont
  {Martinez}(1998)}]{application1}%
  \BibitemOpen
  \bibfield  {author} {\bibinfo {author} {\bibfnamefont {O.}~\bibnamefont
  {Kahn}}\ and\ \bibinfo {author} {\bibfnamefont {C.~J.}\ \bibnamefont
  {Martinez}},\ }\href@noop {} {\bibfield  {journal} {\bibinfo  {journal}
  {Science}\ }\textbf {\bibinfo {volume} {279}},\ \bibinfo {pages} {44}
  (\bibinfo {year} {1998})}\BibitemShut {NoStop}%
\bibitem [{\citenamefont {Moln{\'a}r}\ \emph {et~al.}(2014)\citenamefont
  {Moln{\'a}r}, \citenamefont {Salmon}, \citenamefont {Nicolazzi},
  \citenamefont {Terki},\ and\ \citenamefont {Bousseksou}}]{application2}%
  \BibitemOpen
  \bibfield  {author} {\bibinfo {author} {\bibfnamefont {G.}~\bibnamefont
  {Moln{\'a}r}}, \bibinfo {author} {\bibfnamefont {L.}~\bibnamefont {Salmon}},
  \bibinfo {author} {\bibfnamefont {W.}~\bibnamefont {Nicolazzi}}, \bibinfo
  {author} {\bibfnamefont {F.}~\bibnamefont {Terki}}, \ and\ \bibinfo {author}
  {\bibfnamefont {A.}~\bibnamefont {Bousseksou}},\ }\href@noop {} {\bibfield
  {journal} {\bibinfo  {journal} {Journal of Materials Chemistry C}\ }\textbf
  {\bibinfo {volume} {2}},\ \bibinfo {pages} {1360} (\bibinfo {year}
  {2014})}\BibitemShut {NoStop}%
\bibitem [{\citenamefont {Ruiz}(2014)}]{eliseoruiz2014}%
  \BibitemOpen
  \bibfield  {author} {\bibinfo {author} {\bibfnamefont {E.}~\bibnamefont
  {Ruiz}},\ }\href@noop {} {\bibfield  {journal} {\bibinfo  {journal} {Physical
  Chemistry Chemical Physics}\ }\textbf {\bibinfo {volume} {16}},\ \bibinfo
  {pages} {14} (\bibinfo {year} {2014})}\BibitemShut {NoStop}%
\bibitem [{\citenamefont {G{\"u}tlich}\ and\ \citenamefont
  {Goodwin}(2004)}]{gutlich2004spin}%
  \BibitemOpen
  \bibfield  {author} {\bibinfo {author} {\bibfnamefont {P.}~\bibnamefont
  {G{\"u}tlich}}\ and\ \bibinfo {author} {\bibfnamefont {H.~A.}\ \bibnamefont
  {Goodwin}},\ }\href@noop {} {\bibfield  {journal} {\bibinfo  {journal}
  {Topics in Current Chemistry}\ }\textbf {\bibinfo {volume} {233}},\ \bibinfo
  {pages} {1} (\bibinfo {year} {2004})}\BibitemShut {NoStop}%
\bibitem [{\citenamefont {Chergui}(2013)}]{chergui}%
  \BibitemOpen
  \bibfield  {author} {\bibinfo {author} {\bibfnamefont {M.}~\bibnamefont
  {Chergui}},\ }\enquote {\bibinfo {title} {Ultrafast studies of the
  light-induced spin change in fe(ii)-polypyridine complexes},}\ in\ \href
  {\doibase 10.1002/9781118519301.ch15} {\emph {\bibinfo {booktitle}
  {Spin-Crossover Materials}}}\ (\bibinfo  {publisher} {John Wiley \& Sons
  Ltd},\ \bibinfo {year} {2013})\ pp.\ \bibinfo {pages} {405--424}\BibitemShut
  {NoStop}%
\bibitem [{\citenamefont {Cammarata}\ \emph {et~al.}(2014)\citenamefont
  {Cammarata}, \citenamefont {Bertoni}, \citenamefont {Lorenc}, \citenamefont
  {Cailleau}, \citenamefont {Di~Matteo}, \citenamefont {Mauriac}, \citenamefont
  {Matar}, \citenamefont {Lemke}, \citenamefont {Chollet}, \citenamefont {Ravy}
  \emph {et~al.}}]{cammarata2014prl}%
  \BibitemOpen
  \bibfield  {author} {\bibinfo {author} {\bibfnamefont {M.}~\bibnamefont
  {Cammarata}}, \bibinfo {author} {\bibfnamefont {R.}~\bibnamefont {Bertoni}},
  \bibinfo {author} {\bibfnamefont {M.}~\bibnamefont {Lorenc}}, \bibinfo
  {author} {\bibfnamefont {H.}~\bibnamefont {Cailleau}}, \bibinfo {author}
  {\bibfnamefont {S.}~\bibnamefont {Di~Matteo}}, \bibinfo {author}
  {\bibfnamefont {C.}~\bibnamefont {Mauriac}}, \bibinfo {author} {\bibfnamefont
  {S.~F.}\ \bibnamefont {Matar}}, \bibinfo {author} {\bibfnamefont
  {H.}~\bibnamefont {Lemke}}, \bibinfo {author} {\bibfnamefont
  {M.}~\bibnamefont {Chollet}}, \bibinfo {author} {\bibfnamefont
  {S.}~\bibnamefont {Ravy}},  \emph {et~al.},\ }\href@noop {} {\bibfield
  {journal} {\bibinfo  {journal} {Physical Review Letters}\ }\textbf {\bibinfo
  {volume} {113}},\ \bibinfo {pages} {227402} (\bibinfo {year}
  {2014})}\BibitemShut {NoStop}%
\bibitem [{\citenamefont {Vank{\'o}}\ \emph {et~al.}(2006)\citenamefont
  {Vank{\'o}}, \citenamefont {Neisius}, \citenamefont {Moln{\'a}r},
  \citenamefont {Renz}, \citenamefont {K{\'a}rp{\'a}ti}, \citenamefont
  {Shukla},\ and\ \citenamefont {de~Groot}}]{vanko2006jpcb}%
  \BibitemOpen
  \bibfield  {author} {\bibinfo {author} {\bibfnamefont {G.}~\bibnamefont
  {Vank{\'o}}}, \bibinfo {author} {\bibfnamefont {T.}~\bibnamefont {Neisius}},
  \bibinfo {author} {\bibfnamefont {G.}~\bibnamefont {Moln{\'a}r}}, \bibinfo
  {author} {\bibfnamefont {F.}~\bibnamefont {Renz}}, \bibinfo {author}
  {\bibfnamefont {S.}~\bibnamefont {K{\'a}rp{\'a}ti}}, \bibinfo {author}
  {\bibfnamefont {A.}~\bibnamefont {Shukla}}, \ and\ \bibinfo {author}
  {\bibfnamefont {F.~M.~F.}\ \bibnamefont {de~Groot}},\ }\href {\doibase
  10.1021/jp0615961} {\bibfield  {journal} {\bibinfo  {journal} {The Journal of
  Physical Chemistry B}\ }\textbf {\bibinfo {volume} {110}},\ \bibinfo {pages}
  {11647} (\bibinfo {year} {2006})}\BibitemShut {NoStop}%
\bibitem [{\citenamefont {Canton}\ \emph {et~al.}(2014)\citenamefont {Canton},
  \citenamefont {Zhang}, \citenamefont {Lawson~Daku}, \citenamefont {Smeigh},
  \citenamefont {Zhang}, \citenamefont {Liu}, \citenamefont {Wallentin},
  \citenamefont {Attenkofer}, \citenamefont {Jennings}, \citenamefont {Kurtz}
  \emph {et~al.}}]{canton2014ani}%
  \BibitemOpen
  \bibfield  {author} {\bibinfo {author} {\bibfnamefont {S.~E.}\ \bibnamefont
  {Canton}}, \bibinfo {author} {\bibfnamefont {X.}~\bibnamefont {Zhang}},
  \bibinfo {author} {\bibfnamefont {L.~M.}\ \bibnamefont {Lawson~Daku}},
  \bibinfo {author} {\bibfnamefont {A.~L.}\ \bibnamefont {Smeigh}}, \bibinfo
  {author} {\bibfnamefont {J.}~\bibnamefont {Zhang}}, \bibinfo {author}
  {\bibfnamefont {Y.}~\bibnamefont {Liu}}, \bibinfo {author} {\bibfnamefont
  {C.-J.}\ \bibnamefont {Wallentin}}, \bibinfo {author} {\bibfnamefont
  {K.}~\bibnamefont {Attenkofer}}, \bibinfo {author} {\bibfnamefont
  {G.}~\bibnamefont {Jennings}}, \bibinfo {author} {\bibfnamefont {C.~A.}\
  \bibnamefont {Kurtz}},  \emph {et~al.},\ }\href@noop {} {\bibfield  {journal}
  {\bibinfo  {journal} {The Journal of Physical Chemistry C}\ }\textbf
  {\bibinfo {volume} {118}},\ \bibinfo {pages} {4536} (\bibinfo {year}
  {2014})}\BibitemShut {NoStop}%
\bibitem [{\citenamefont {Nance}\ \emph {et~al.}(2015)\citenamefont {Nance},
  \citenamefont {Bowman}, \citenamefont {Mukherjee}, \citenamefont {Kelley},\
  and\ \citenamefont {Jakubikova}}]{Nance2015}%
  \BibitemOpen
  \bibfield  {author} {\bibinfo {author} {\bibfnamefont {J.}~\bibnamefont
  {Nance}}, \bibinfo {author} {\bibfnamefont {D.~N.}\ \bibnamefont {Bowman}},
  \bibinfo {author} {\bibfnamefont {S.}~\bibnamefont {Mukherjee}}, \bibinfo
  {author} {\bibfnamefont {C.~T.}\ \bibnamefont {Kelley}}, \ and\ \bibinfo
  {author} {\bibfnamefont {E.}~\bibnamefont {Jakubikova}},\ }\href@noop {}
  {\bibfield  {journal} {\bibinfo  {journal} {Inorganic Chemistry}\ }\textbf
  {\bibinfo {volume} {54}},\ \bibinfo {pages} {11259–11268} (\bibinfo {year}
  {2015})}\BibitemShut {NoStop}%
\bibitem [{\citenamefont {Haldrup}\ \emph {et~al.}(2016)\citenamefont
  {Haldrup}, \citenamefont {Gawelda}, \citenamefont {Abela}, \citenamefont
  {Alonso-Mori}, \citenamefont {Bergmann}, \citenamefont {Bordage},
  \citenamefont {Cammarata}, \citenamefont {Canton}, \citenamefont {Dohn},
  \citenamefont {van Driel}, \citenamefont {Fritz}, \citenamefont {Galler},
  \citenamefont {Glatzel}, \citenamefont {Harlang}, \citenamefont {Kjær},
  \citenamefont {Lemke}, \citenamefont {Moller}, \citenamefont {Németh},
  \citenamefont {Pápai}, \citenamefont {Sas}, \citenamefont {Uhlig},
  \citenamefont {Zhu}, \citenamefont {Vankó}, \citenamefont {Sundstrom},
  \citenamefont {Nielsen},\ and\ \citenamefont
  {Bressler}}]{Haldrup2016_LCLS2011}%
  \BibitemOpen
  \bibfield  {author} {\bibinfo {author} {\bibfnamefont {K.}~\bibnamefont
  {Haldrup}}, \bibinfo {author} {\bibfnamefont {W.}~\bibnamefont {Gawelda}},
  \bibinfo {author} {\bibfnamefont {R.}~\bibnamefont {Abela}}, \bibinfo
  {author} {\bibfnamefont {R.}~\bibnamefont {Alonso-Mori}}, \bibinfo {author}
  {\bibfnamefont {U.}~\bibnamefont {Bergmann}}, \bibinfo {author}
  {\bibfnamefont {A.}~\bibnamefont {Bordage}}, \bibinfo {author} {\bibfnamefont
  {M.}~\bibnamefont {Cammarata}}, \bibinfo {author} {\bibfnamefont
  {S.}~\bibnamefont {Canton}}, \bibinfo {author} {\bibfnamefont
  {A.}~\bibnamefont {Dohn}}, \bibinfo {author} {\bibfnamefont {T.}~\bibnamefont
  {van Driel}}, \bibinfo {author} {\bibfnamefont {D.}~\bibnamefont {Fritz}},
  \bibinfo {author} {\bibfnamefont {A.}~\bibnamefont {Galler}}, \bibinfo
  {author} {\bibfnamefont {P.}~\bibnamefont {Glatzel}}, \bibinfo {author}
  {\bibfnamefont {T.}~\bibnamefont {Harlang}}, \bibinfo {author} {\bibfnamefont
  {K.}~\bibnamefont {Kjær}}, \bibinfo {author} {\bibfnamefont
  {H.}~\bibnamefont {Lemke}}, \bibinfo {author} {\bibfnamefont
  {K.}~\bibnamefont {Moller}}, \bibinfo {author} {\bibfnamefont
  {Z.}~\bibnamefont {Németh}}, \bibinfo {author} {\bibfnamefont
  {M.}~\bibnamefont {Pápai}}, \bibinfo {author} {\bibfnamefont
  {N.}~\bibnamefont {Sas}}, \bibinfo {author} {\bibfnamefont {J.}~\bibnamefont
  {Uhlig}}, \bibinfo {author} {\bibfnamefont {D.}~\bibnamefont {Zhu}}, \bibinfo
  {author} {\bibfnamefont {G.}~\bibnamefont {Vankó}}, \bibinfo {author}
  {\bibfnamefont {V.}~\bibnamefont {Sundstrom}}, \bibinfo {author}
  {\bibfnamefont {M.}~\bibnamefont {Nielsen}}, \ and\ \bibinfo {author}
  {\bibfnamefont {C.}~\bibnamefont {Bressler}},\ }\href@noop {} {\bibfield
  {journal} {\bibinfo  {journal} {The Journal of Physical Chemistry B}\
  }\textbf {\bibinfo {volume} {120}},\ \bibinfo {pages} {1158} (\bibinfo {year}
  {2016})}\BibitemShut {NoStop}%
\bibitem [{\citenamefont {Lemke}()}]{lemkeAr}%
  \BibitemOpen
  \bibfield  {author} {\bibinfo {author} {\bibfnamefont {H.}\
  \bibnamefont {Lemke et al.}},\ }\href@noop {} {\ }\bibinfo {note} {ArXiv:
  1511.01294}\BibitemShut {NoStop}%
\bibitem [{\citenamefont {Lemke}\ \emph {et~al.}(2013)\citenamefont {Lemke},
  \citenamefont {Bressler}, \citenamefont {Chen}, \citenamefont {Fritz},
  \citenamefont {Gaffney}, \citenamefont {Galler}, \citenamefont {Gawelda},
  \citenamefont {Haldrup}, \citenamefont {Hartsock}, \citenamefont {Ihee},
  \citenamefont {Kim}, \citenamefont {Kim}, \citenamefont {Lee}, \citenamefont
  {Nielsen}, \citenamefont {Stickrath}, \citenamefont {Zhang}, \citenamefont
  {Zhu},\ and\ \citenamefont {Cammarata}}]{lemke}%
  \BibitemOpen
  \bibfield  {author} {\bibinfo {author} {\bibfnamefont {H.~T.}\ \bibnamefont
  {Lemke}}, \bibinfo {author} {\bibfnamefont {C.}~\bibnamefont {Bressler}},
  \bibinfo {author} {\bibfnamefont {L.~X.}\ \bibnamefont {Chen}}, \bibinfo
  {author} {\bibfnamefont {D.~M.}\ \bibnamefont {Fritz}}, \bibinfo {author}
  {\bibfnamefont {K.~J.}\ \bibnamefont {Gaffney}}, \bibinfo {author}
  {\bibfnamefont {A.}~\bibnamefont {Galler}}, \bibinfo {author} {\bibfnamefont
  {W.}~\bibnamefont {Gawelda}}, \bibinfo {author} {\bibfnamefont
  {K.}~\bibnamefont {Haldrup}}, \bibinfo {author} {\bibfnamefont {R.~W.}\
  \bibnamefont {Hartsock}}, \bibinfo {author} {\bibfnamefont {H.}~\bibnamefont
  {Ihee}}, \bibinfo {author} {\bibfnamefont {J.}~\bibnamefont {Kim}}, \bibinfo
  {author} {\bibfnamefont {K.~H.}\ \bibnamefont {Kim}}, \bibinfo {author}
  {\bibfnamefont {J.~H.}\ \bibnamefont {Lee}}, \bibinfo {author} {\bibfnamefont
  {M.~M.}\ \bibnamefont {Nielsen}}, \bibinfo {author} {\bibfnamefont {A.~B.}\
  \bibnamefont {Stickrath}}, \bibinfo {author} {\bibfnamefont {W.}~\bibnamefont
  {Zhang}}, \bibinfo {author} {\bibfnamefont {D.}~\bibnamefont {Zhu}}, \ and\
  \bibinfo {author} {\bibfnamefont {M.}~\bibnamefont {Cammarata}},\ }\href
  {\doibase 10.1021/jp312559h} {\bibfield  {journal} {\bibinfo  {journal} {The
  Journal of Physical Chemistry A}\ }\textbf {\bibinfo {volume} {117}},\
  \bibinfo {pages} {735–740} (\bibinfo {year} {2013})}\BibitemShut {NoStop}%
\bibitem [{\citenamefont {Chollet}\ \emph {et~al.}(2015)\citenamefont
  {Chollet}, \citenamefont {Alonso-Mori}, \citenamefont {Cammarata},
  \citenamefont {Damiani}, \citenamefont {Defever}, \citenamefont {Delor},
  \citenamefont {Feng}, \citenamefont {Glownia}, \citenamefont {Langton},
  \citenamefont {Nelson}, \citenamefont {Ramsey}, \citenamefont {Robert},
  \citenamefont {Sikorski}, \citenamefont {Song}, \citenamefont {Stefanescu},
  \citenamefont {Srinivasan}, \citenamefont {Zhu}, \citenamefont {Lemke},\ and\
  \citenamefont {Fritz}}]{xpp}%
  \BibitemOpen
  \bibfield  {author} {\bibinfo {author} {\bibfnamefont {M.}~\bibnamefont
  {Chollet}}, \bibinfo {author} {\bibfnamefont {R.}~\bibnamefont
  {Alonso-Mori}}, \bibinfo {author} {\bibfnamefont {M.}~\bibnamefont
  {Cammarata}}, \bibinfo {author} {\bibfnamefont {D.}~\bibnamefont {Damiani}},
  \bibinfo {author} {\bibfnamefont {J.}~\bibnamefont {Defever}}, \bibinfo
  {author} {\bibfnamefont {J.~T.}\ \bibnamefont {Delor}}, \bibinfo {author}
  {\bibfnamefont {Y.}~\bibnamefont {Feng}}, \bibinfo {author} {\bibfnamefont
  {J.~M.}\ \bibnamefont {Glownia}}, \bibinfo {author} {\bibfnamefont {J.~B.}\
  \bibnamefont {Langton}}, \bibinfo {author} {\bibfnamefont {S.}~\bibnamefont
  {Nelson}}, \bibinfo {author} {\bibfnamefont {K.}~\bibnamefont {Ramsey}},
  \bibinfo {author} {\bibfnamefont {A.}~\bibnamefont {Robert}}, \bibinfo
  {author} {\bibfnamefont {M.}~\bibnamefont {Sikorski}}, \bibinfo {author}
  {\bibfnamefont {S.}~\bibnamefont {Song}}, \bibinfo {author} {\bibfnamefont
  {D.}~\bibnamefont {Stefanescu}}, \bibinfo {author} {\bibfnamefont
  {V.}~\bibnamefont {Srinivasan}}, \bibinfo {author} {\bibfnamefont
  {D.}~\bibnamefont {Zhu}}, \bibinfo {author} {\bibfnamefont {H.~T.}\
  \bibnamefont {Lemke}}, \ and\ \bibinfo {author} {\bibfnamefont {D.~M.}\
  \bibnamefont {Fritz}},\ }\href@noop {} {\bibfield  {journal} {\bibinfo
  {journal} {Journal of Synchrotron Radiation}\ }\textbf {\bibinfo {volume}
  {22}},\ \bibinfo {pages} {503} (\bibinfo {year} {2015})}\BibitemShut
  {NoStop}%
\bibitem [{\citenamefont {Kremer}\ \emph {et~al.}(1982)\citenamefont {Kremer},
  \citenamefont {Henke},\ and\ \citenamefont {Reinen}}]{Kremer}%
  \BibitemOpen
  \bibfield  {author} {\bibinfo {author} {\bibfnamefont {S.}~\bibnamefont
  {Kremer}}, \bibinfo {author} {\bibfnamefont {W.}~\bibnamefont {Henke}}, \
  and\ \bibinfo {author} {\bibfnamefont {D.}~\bibnamefont {Reinen}},\
  }\href@noop {} {\bibfield  {journal} {\bibinfo  {journal} {Inorganic
  Chemistry}\ }\textbf {\bibinfo {volume} {21}},\ \bibinfo {pages}
  {3013–3022} (\bibinfo {year} {1982})}\BibitemShut {NoStop}%
\bibitem [{\citenamefont {Judge}\ and\ \citenamefont
  {Baker~Jr.}(1967)}]{JudgeBaker1967}%
  \BibitemOpen
  \bibfield  {author} {\bibinfo {author} {\bibfnamefont {J.~S.}\ \bibnamefont
  {Judge}}\ and\ \bibinfo {author} {\bibfnamefont {W.}~\bibnamefont
  {Baker~Jr.}},\ }\href@noop {} {\bibfield  {journal} {\bibinfo  {journal}
  {Inorganica Chimica Acta}\ }\textbf {\bibinfo {volume} {1}},\ \bibinfo
  {pages} {68} (\bibinfo {year} {1967})}\BibitemShut {NoStop}%
\bibitem [{\citenamefont {Hayami}\ \emph {et~al.}(2011)\citenamefont {Hayami},
  \citenamefont {Komatsu}, \citenamefont {Shimizu}, \citenamefont {Kamihata},\
  and\ \citenamefont {Lee}}]{hayami2011SCOterpy}%
  \BibitemOpen
  \bibfield  {author} {\bibinfo {author} {\bibfnamefont {S.}~\bibnamefont
  {Hayami}}, \bibinfo {author} {\bibfnamefont {Y.}~\bibnamefont {Komatsu}},
  \bibinfo {author} {\bibfnamefont {T.}~\bibnamefont {Shimizu}}, \bibinfo
  {author} {\bibfnamefont {H.}~\bibnamefont {Kamihata}}, \ and\ \bibinfo
  {author} {\bibfnamefont {Y.~H.}\ \bibnamefont {Lee}},\ }\href@noop {}
  {\bibfield  {journal} {\bibinfo  {journal} {Coordination Chemistry Reviews}\
  }\textbf {\bibinfo {volume} {255}},\ \bibinfo {pages} {1981} (\bibinfo {year}
  {2011})}\BibitemShut {NoStop}%
\bibitem [{\citenamefont {Oshio}\ \emph {et~al.}(2001)\citenamefont {Oshio},
  \citenamefont {Spiering}, \citenamefont {Ksenofontov}, \citenamefont {Renz},\
  and\ \citenamefont {G{\"u}tlich}}]{oshio}%
  \BibitemOpen
  \bibfield  {author} {\bibinfo {author} {\bibfnamefont {H.}~\bibnamefont
  {Oshio}}, \bibinfo {author} {\bibfnamefont {H.}~\bibnamefont {Spiering}},
  \bibinfo {author} {\bibfnamefont {V.}~\bibnamefont {Ksenofontov}}, \bibinfo
  {author} {\bibfnamefont {F.}~\bibnamefont {Renz}}, \ and\ \bibinfo {author}
  {\bibfnamefont {P.}~\bibnamefont {G{\"u}tlich}},\ }\href@noop {} {\bibfield
  {journal} {\bibinfo  {journal} {Inorganic Chemistry}\ }\textbf {\bibinfo
  {volume} {40}},\ \bibinfo {pages} {1143} (\bibinfo {year}
  {2001})}\BibitemShut {NoStop}%
\bibitem [{\citenamefont {Vargas}\ \emph {et~al.}(2013)\citenamefont {Vargas},
  \citenamefont {Krivokapic}, \citenamefont {Hauser},\ and\ \citenamefont
  {Lawson~Daku}}]{Vargas}%
  \BibitemOpen
  \bibfield  {author} {\bibinfo {author} {\bibfnamefont {A.}~\bibnamefont
  {Vargas}}, \bibinfo {author} {\bibfnamefont {I.}~\bibnamefont {Krivokapic}},
  \bibinfo {author} {\bibfnamefont {A.}~\bibnamefont {Hauser}}, \ and\ \bibinfo
  {author} {\bibfnamefont {L.~M.}\ \bibnamefont {Lawson~Daku}},\ }\href@noop {}
  {\bibfield  {journal} {\bibinfo  {journal} {Physical Chemistry Chemical
  Physics}\ }\textbf {\bibinfo {volume} {15}},\ \bibinfo {pages} {3752}
  (\bibinfo {year} {2013})}\BibitemShut {NoStop}%
\bibitem [{\citenamefont {Figgis}\ \emph {et~al.}(1983)\citenamefont {Figgis},
  \citenamefont {Kucharski},\ and\ \citenamefont {White}}]{crystalStructure}%
  \BibitemOpen
  \bibfield  {author} {\bibinfo {author} {\bibfnamefont {B.~N.}\ \bibnamefont
  {Figgis}}, \bibinfo {author} {\bibfnamefont {E.~S.}\ \bibnamefont
  {Kucharski}}, \ and\ \bibinfo {author} {\bibfnamefont {A.~H.}\ \bibnamefont
  {White}},\ }\href@noop {} {\bibfield  {journal} {\bibinfo  {journal}
  {Australian Journal of Chemistry}\ }\textbf {\bibinfo {volume} {36}},\
  \bibinfo {pages} {1537} (\bibinfo {year} {1983})}\BibitemShut {NoStop}%
\bibitem [{\citenamefont {Beattie}\ \emph {et~al.}(1995)\citenamefont
  {Beattie}, \citenamefont {Binstead}, \citenamefont {Kelso}, \citenamefont
  {Del~Favero}, \citenamefont {Dewey},\ and\ \citenamefont
  {Turner}}]{beattie1995dynamics}%
  \BibitemOpen
  \bibfield  {author} {\bibinfo {author} {\bibfnamefont {J.~K.}\ \bibnamefont
  {Beattie}}, \bibinfo {author} {\bibfnamefont {R.~A.}\ \bibnamefont
  {Binstead}}, \bibinfo {author} {\bibfnamefont {M.~T.}\ \bibnamefont {Kelso}},
  \bibinfo {author} {\bibfnamefont {P.}~\bibnamefont {Del~Favero}}, \bibinfo
  {author} {\bibfnamefont {T.~G.}\ \bibnamefont {Dewey}}, \ and\ \bibinfo
  {author} {\bibfnamefont {D.~H.}\ \bibnamefont {Turner}},\ }\href@noop {}
  {\bibfield  {journal} {\bibinfo  {journal} {Inorganica Chimica Acta}\
  }\textbf {\bibinfo {volume} {235}},\ \bibinfo {pages} {245} (\bibinfo {year}
  {1995})}\BibitemShut {NoStop}%
\bibitem [{\citenamefont {Minitti}\ \emph {et~al.}(2015)\citenamefont
  {Minitti}, \citenamefont {Robinson}, \citenamefont {Coffee}, \citenamefont
  {Edstrom}, \citenamefont {Gilevich}, \citenamefont {Glownia}, \citenamefont
  {Granados}, \citenamefont {Hering}, \citenamefont {Hoffmann}, \citenamefont
  {Miahnahri},\ and\ \citenamefont {et~al.}}]{timingtoolXPP}%
  \BibitemOpen
  \bibfield  {author} {\bibinfo {author} {\bibfnamefont {M.~P.}\ \bibnamefont
  {Minitti}}, \bibinfo {author} {\bibfnamefont {J.~S.}\ \bibnamefont
  {Robinson}}, \bibinfo {author} {\bibfnamefont {R.~N.}\ \bibnamefont
  {Coffee}}, \bibinfo {author} {\bibfnamefont {S.}~\bibnamefont {Edstrom}},
  \bibinfo {author} {\bibfnamefont {S.}~\bibnamefont {Gilevich}}, \bibinfo
  {author} {\bibfnamefont {J.~M.}\ \bibnamefont {Glownia}}, \bibinfo {author}
  {\bibfnamefont {E.}~\bibnamefont {Granados}}, \bibinfo {author}
  {\bibfnamefont {P.}~\bibnamefont {Hering}}, \bibinfo {author} {\bibfnamefont
  {M.~C.}\ \bibnamefont {Hoffmann}}, \bibinfo {author} {\bibfnamefont
  {A.}~\bibnamefont {Miahnahri}}, \ and\ \bibinfo {author} {\bibnamefont
  {et~al.}},\ }\href@noop {} {\bibfield  {journal} {\bibinfo  {journal}
  {Journal of Synchrotron Radiation}\ }\textbf {\bibinfo {volume} {22}},\
  \bibinfo {pages} {526} (\bibinfo {year} {2015})}\BibitemShut {NoStop}%
\bibitem [{\citenamefont {Philipp}\ \emph {et~al.}(2011)\citenamefont
  {Philipp}, \citenamefont {Hromalik}, \citenamefont {Tate}, \citenamefont
  {Koerner},\ and\ \citenamefont {Gruner}}]{cspad}%
  \BibitemOpen
  \bibfield  {author} {\bibinfo {author} {\bibfnamefont {H.~T.}\ \bibnamefont
  {Philipp}}, \bibinfo {author} {\bibfnamefont {M.}~\bibnamefont {Hromalik}},
  \bibinfo {author} {\bibfnamefont {M.}~\bibnamefont {Tate}}, \bibinfo {author}
  {\bibfnamefont {L.}~\bibnamefont {Koerner}}, \ and\ \bibinfo {author}
  {\bibfnamefont {S.~M.}\ \bibnamefont {Gruner}},\ }\href@noop {} {\bibfield
  {journal} {\bibinfo  {journal} {Nuclear Instruments and Methods in Physics
  Research Section A}\ }\textbf {\bibinfo {volume} {649}},\ \bibinfo {pages}
  {67} (\bibinfo {year} {2011})}\BibitemShut {NoStop}%
\bibitem [{\citenamefont {van Driel}\ \emph {et~al.}(2015)\citenamefont {van
  Driel}, \citenamefont {Kjaer}, \citenamefont {Biasin}, \citenamefont
  {Haldrup}, \citenamefont {Lemke},\ and\ \citenamefont {Nielsen}}]{timSVD}%
  \BibitemOpen
  \bibfield  {author} {\bibinfo {author} {\bibfnamefont {T.~B.}\ \bibnamefont
  {van Driel}}, \bibinfo {author} {\bibfnamefont {K.~S.}\ \bibnamefont
  {Kjaer}}, \bibinfo {author} {\bibfnamefont {E.}~\bibnamefont {Biasin}},
  \bibinfo {author} {\bibfnamefont {K.}~\bibnamefont {Haldrup}}, \bibinfo
  {author} {\bibfnamefont {H.~T.}\ \bibnamefont {Lemke}}, \ and\ \bibinfo
  {author} {\bibfnamefont {M.~M.}\ \bibnamefont {Nielsen}},\ }\href {\doibase
  10.1039/C4FD00203B} {\bibfield  {journal} {\bibinfo  {journal} {Faraday
  Discussions}\ }\textbf {\bibinfo {volume} {177}},\ \bibinfo {pages} {443}
  (\bibinfo {year} {2015})}\BibitemShut {NoStop}%
\bibitem [{\citenamefont {Haldrup}\ \emph {et~al.}(2010)\citenamefont
  {Haldrup}, \citenamefont {Christensen},\ and\ \citenamefont
  {Meedom~Nielsen}}]{haldrup2010analysis}%
  \BibitemOpen
  \bibfield  {author} {\bibinfo {author} {\bibfnamefont {K.}~\bibnamefont
  {Haldrup}}, \bibinfo {author} {\bibfnamefont {M.}~\bibnamefont
  {Christensen}}, \ and\ \bibinfo {author} {\bibfnamefont {M.}~\bibnamefont
  {Meedom~Nielsen}},\ }\href@noop {} {\bibfield  {journal} {\bibinfo  {journal}
  {Acta Crystallographica Section A}\ }\textbf {\bibinfo {volume} {66}},\
  \bibinfo {pages} {261} (\bibinfo {year} {2010})}\BibitemShut {NoStop}%
\bibitem [{\citenamefont {Lorenz}\ \emph {et~al.}(2010)\citenamefont {Lorenz},
  \citenamefont {Møller},\ and\ \citenamefont {Henriksen}}]{aniUlf}%
  \BibitemOpen
  \bibfield  {author} {\bibinfo {author} {\bibfnamefont {U.}~\bibnamefont
  {Lorenz}}, \bibinfo {author} {\bibfnamefont {K.~B.}\ \bibnamefont {Møller}},
  \ and\ \bibinfo {author} {\bibfnamefont {N.~E.}\ \bibnamefont {Henriksen}},\
  }\href@noop {} {\bibfield  {journal} {\bibinfo  {journal} {New Journal of
  Physics}\ }\textbf {\bibinfo {volume} {12}},\ \bibinfo {pages} {113022}
  (\bibinfo {year} {2010})}\BibitemShut {NoStop}%
\bibitem [{\citenamefont {Haldrup}\ \emph {et~al.}(2012)\citenamefont
  {Haldrup}, \citenamefont {Vankó}, \citenamefont {Gawelda}, \citenamefont
  {Galler}, \citenamefont {Doumy}, \citenamefont {March}, \citenamefont
  {Kanter}, \citenamefont {Bordage}, \citenamefont {Dohn}, \citenamefont {van
  Driel}, \citenamefont {Kjær}, \citenamefont {Lemke}, \citenamefont {Canton},
  \citenamefont {Uhlig}, \citenamefont {Sundström}, \citenamefont {Young},
  \citenamefont {Southworth}, \citenamefont {Nielsen},\ and\ \citenamefont
  {Bressler}}]{Kreuff-Fe}%
  \BibitemOpen
  \bibfield  {author} {\bibinfo {author} {\bibfnamefont {K.}~\bibnamefont
  {Haldrup}}, \bibinfo {author} {\bibfnamefont {G.}~\bibnamefont {Vankó}},
  \bibinfo {author} {\bibfnamefont {W.}~\bibnamefont {Gawelda}}, \bibinfo
  {author} {\bibfnamefont {A.}~\bibnamefont {Galler}}, \bibinfo {author}
  {\bibfnamefont {G.}~\bibnamefont {Doumy}}, \bibinfo {author} {\bibfnamefont
  {A.~M.}\ \bibnamefont {March}}, \bibinfo {author} {\bibfnamefont {E.~P.}\
  \bibnamefont {Kanter}}, \bibinfo {author} {\bibfnamefont {A.}~\bibnamefont
  {Bordage}}, \bibinfo {author} {\bibfnamefont {A.}~\bibnamefont {Dohn}},
  \bibinfo {author} {\bibfnamefont {T.~B.}\ \bibnamefont {van Driel}}, \bibinfo
  {author} {\bibfnamefont {K.~S.}\ \bibnamefont {Kjær}}, \bibinfo {author}
  {\bibfnamefont {H.~T.}\ \bibnamefont {Lemke}}, \bibinfo {author}
  {\bibfnamefont {S.~E.}\ \bibnamefont {Canton}}, \bibinfo {author}
  {\bibfnamefont {J.}~\bibnamefont {Uhlig}}, \bibinfo {author} {\bibfnamefont
  {V.}~\bibnamefont {Sundström}}, \bibinfo {author} {\bibfnamefont
  {L.}~\bibnamefont {Young}}, \bibinfo {author} {\bibfnamefont {S.~H.}\
  \bibnamefont {Southworth}}, \bibinfo {author} {\bibfnamefont {M.~M.}\
  \bibnamefont {Nielsen}}, \ and\ \bibinfo {author} {\bibfnamefont
  {C.}~\bibnamefont {Bressler}},\ }\href@noop {} {\bibfield  {journal}
  {\bibinfo  {journal} {The Journal of Physical Chemistry A}\ }\textbf
  {\bibinfo {volume} {116}},\ \bibinfo {pages} {9878} (\bibinfo {year}
  {2012})}\BibitemShut {NoStop}%
\bibitem [{sm()}]{sm}%
  \BibitemOpen
  \href@noop {} {\ }\bibinfo {note} {See Supplemental Material at
  http://dx.doi.org/10.1103/PhysRevLett.117.013002 for the full description of
  the fit procedure, a further analysis and discussion of the solvent and
  structural contribution to the difference scattering signal as a function of
  laser power and an extended description of the dynamics observed in the data,
  including movies S1 and S2 showing the identified vibrational modes. Details
  of the IRF-determination and MD simulations are also included.}\BibitemShut
  {Stop}%
\bibitem [{\citenamefont {Dohn}\ \emph {et~al.}(2015)\citenamefont {Dohn},
  \citenamefont {Biasin}, \citenamefont {Haldrup}, \citenamefont {Nielsen},
  \citenamefont {Henriksen},\ and\ \citenamefont {Møller}}]{asmusRDF}%
  \BibitemOpen
  \bibfield  {author} {\bibinfo {author} {\bibfnamefont {A.~O.}\ \bibnamefont
  {Dohn}}, \bibinfo {author} {\bibfnamefont {E.}~\bibnamefont {Biasin}},
  \bibinfo {author} {\bibfnamefont {K.}~\bibnamefont {Haldrup}}, \bibinfo
  {author} {\bibfnamefont {M.~M.}\ \bibnamefont {Nielsen}}, \bibinfo {author}
  {\bibfnamefont {N.~E.}\ \bibnamefont {Henriksen}}, \ and\ \bibinfo {author}
  {\bibfnamefont {K.~B.}\ \bibnamefont {Møller}},\ }\href {\doibase
  10.1088/0953-4075/48/24/244010} {\bibfield  {journal} {\bibinfo  {journal}
  {ournal of Physics B: Atomic, Molecular and Optical Physics}\ }\textbf
  {\bibinfo {volume} {48}},\ \bibinfo {pages} {244010} (\bibinfo {year}
  {2015})}\BibitemShut {NoStop}%
\bibitem [{\citenamefont {Cammarata}\ \emph {et~al.}(2006)\citenamefont
  {Cammarata}, \citenamefont {Lorenc}, \citenamefont {Kim}, \citenamefont
  {Lee}, \citenamefont {Kong}, \citenamefont {Pontecorvo}, \citenamefont
  {Russo}, \citenamefont {Schiro}, \citenamefont {Cupane}, \citenamefont
  {Wulff} \emph {et~al.}}]{cammarata2006solvent}%
  \BibitemOpen
  \bibfield  {author} {\bibinfo {author} {\bibfnamefont {M.}~\bibnamefont
  {Cammarata}}, \bibinfo {author} {\bibfnamefont {M.}~\bibnamefont {Lorenc}},
  \bibinfo {author} {\bibfnamefont {T.}~\bibnamefont {Kim}}, \bibinfo {author}
  {\bibfnamefont {J.}~\bibnamefont {Lee}}, \bibinfo {author} {\bibfnamefont
  {Q.}~\bibnamefont {Kong}}, \bibinfo {author} {\bibfnamefont {E.}~\bibnamefont
  {Pontecorvo}}, \bibinfo {author} {\bibfnamefont {M.~L.}\ \bibnamefont
  {Russo}}, \bibinfo {author} {\bibfnamefont {G.}~\bibnamefont {Schiro}},
  \bibinfo {author} {\bibfnamefont {A.}~\bibnamefont {Cupane}}, \bibinfo
  {author} {\bibfnamefont {M.}~\bibnamefont {Wulff}},  \emph {et~al.},\
  }\href@noop {} {\bibfield  {journal} {\bibinfo  {journal} {The Journal of
  Chemical Physics}\ }\textbf {\bibinfo {volume} {124}},\ \bibinfo {pages}
  {124504} (\bibinfo {year} {2006})}\BibitemShut {NoStop}%
\bibitem [{\citenamefont {Kjaer}\ \emph {et~al.}(2013)\citenamefont {Kjaer},
  \citenamefont {van Driel}, \citenamefont {Kehres}, \citenamefont {Haldrup},
  \citenamefont {Khakhulin}, \citenamefont {Bechgaard}, \citenamefont
  {Cammarata}, \citenamefont {Wulff}, \citenamefont {Sørensen},\ and\
  \citenamefont {Nielsen}}]{KasperSolvent}%
  \BibitemOpen
  \bibfield  {author} {\bibinfo {author} {\bibfnamefont {K.~S.}\ \bibnamefont
  {Kjaer}}, \bibinfo {author} {\bibfnamefont {T.~B.}\ \bibnamefont {van
  Driel}}, \bibinfo {author} {\bibfnamefont {J.}~\bibnamefont {Kehres}},
  \bibinfo {author} {\bibfnamefont {K.}~\bibnamefont {Haldrup}}, \bibinfo
  {author} {\bibfnamefont {D.}~\bibnamefont {Khakhulin}}, \bibinfo {author}
  {\bibfnamefont {K.}~\bibnamefont {Bechgaard}}, \bibinfo {author}
  {\bibfnamefont {M.}~\bibnamefont {Cammarata}}, \bibinfo {author}
  {\bibfnamefont {M.}~\bibnamefont {Wulff}}, \bibinfo {author} {\bibfnamefont
  {T.~J.}\ \bibnamefont {Sørensen}}, \ and\ \bibinfo {author} {\bibfnamefont
  {M.~M.}\ \bibnamefont {Nielsen}},\ }\href {\doibase 10.1039/C3CP50751C}
  {\bibfield  {journal} {\bibinfo  {journal} {Physical Chemistry Chemical
  Physics}\ }\textbf {\bibinfo {volume} {15}},\ \bibinfo {pages} {15003}
  (\bibinfo {year} {2013})}\BibitemShut {NoStop}%
\bibitem [{\citenamefont {Sørensen}\ and\ \citenamefont
  {Kjær}(2013)}]{URLsolvent}%
  \BibitemOpen
  \bibfield  {author} {\bibinfo {author} {\bibfnamefont {T.~J.}\ \bibnamefont
  {Sørensen}}\ and\ \bibinfo {author} {\bibfnamefont {K.~S.}\ \bibnamefont
  {Kjær}},\ }\href@noop {} {} (\bibinfo {year} {2013}),\ \bibinfo {note}
  {https://sites.google.com/site/trwaxs/}\BibitemShut {NoStop}%
\bibitem [{\citenamefont {Jun}\ \emph {et~al.}(2010)\citenamefont {Jun},
  \citenamefont {Lee}, \citenamefont {Kim}, \citenamefont {Kim}, \citenamefont
  {Kim}, \citenamefont {Kong}, \citenamefont {Kim}, \citenamefont {Lo~Russo},
  \citenamefont {Wulff},\ and\ \citenamefont {Ihee}}]{Jun_Ihee_2010}%
  \BibitemOpen
  \bibfield  {author} {\bibinfo {author} {\bibfnamefont {S.}~\bibnamefont
  {Jun}}, \bibinfo {author} {\bibfnamefont {J.~H.}\ \bibnamefont {Lee}},
  \bibinfo {author} {\bibfnamefont {J.}~\bibnamefont {Kim}}, \bibinfo {author}
  {\bibfnamefont {J.}~\bibnamefont {Kim}}, \bibinfo {author} {\bibfnamefont
  {K.~H.}\ \bibnamefont {Kim}}, \bibinfo {author} {\bibfnamefont
  {Q.}~\bibnamefont {Kong}}, \bibinfo {author} {\bibfnamefont {T.~K.}\
  \bibnamefont {Kim}}, \bibinfo {author} {\bibfnamefont {M.}~\bibnamefont
  {Lo~Russo}}, \bibinfo {author} {\bibfnamefont {M.}~\bibnamefont {Wulff}}, \
  and\ \bibinfo {author} {\bibfnamefont {H.}~\bibnamefont {Ihee}},\ }\href
  {\doibase 10.1039/C002004D} {\bibfield  {journal} {\bibinfo  {journal}
  {Physical Chemistry Chemical Physics}\ }\textbf {\bibinfo {volume} {12}},\
  \bibinfo {pages} {11536} (\bibinfo {year} {2010})}\BibitemShut {NoStop}%
\bibitem [{\citenamefont {Auböck}\ and\ \citenamefont
  {Chergui}(2015)}]{Majed-FeBpy2015}%
  \BibitemOpen
  \bibfield  {author} {\bibinfo {author} {\bibfnamefont {G.}~\bibnamefont
  {Auböck}}\ and\ \bibinfo {author} {\bibfnamefont {M.}~\bibnamefont
  {Chergui}},\ }\href@noop {} {\bibfield  {journal} {\bibinfo  {journal}
  {Nature Chemistry}\ }\textbf {\bibinfo {volume} {7}},\ \bibinfo {pages} {629}
  (\bibinfo {year} {2015})}\BibitemShut {NoStop}%
\bibitem [{\citenamefont {Consani}\ \emph {et~al.}(2009)\citenamefont
  {Consani}, \citenamefont {Prémont-Schwarz}, \citenamefont {ElNahhas},
  \citenamefont {Bressler}, \citenamefont {van Mourik}, \citenamefont
  {Cannizzo},\ and\ \citenamefont {Chergui}}]{consani2009}%
  \BibitemOpen
  \bibfield  {author} {\bibinfo {author} {\bibfnamefont {C.}~\bibnamefont
  {Consani}}, \bibinfo {author} {\bibfnamefont {M.}~\bibnamefont
  {Prémont-Schwarz}}, \bibinfo {author} {\bibfnamefont {A.}~\bibnamefont
  {ElNahhas}}, \bibinfo {author} {\bibfnamefont {C.}~\bibnamefont {Bressler}},
  \bibinfo {author} {\bibfnamefont {F.}~\bibnamefont {van Mourik}}, \bibinfo
  {author} {\bibfnamefont {A.}~\bibnamefont {Cannizzo}}, \ and\ \bibinfo
  {author} {\bibfnamefont {M.}~\bibnamefont {Chergui}},\ }\href {\doibase
  10.1002/ange.200902728} {\bibfield  {journal} {\bibinfo  {journal}
  {Angewandte Chemie}\ }\textbf {\bibinfo {volume} {121}},\ \bibinfo {pages}
  {7320} (\bibinfo {year} {2009})}\BibitemShut {NoStop}%
\bibitem [{\citenamefont {Vankó}\ \emph {et~al.}(2015)\citenamefont {Vankó},
  \citenamefont {Bordage}, \citenamefont {Pápai}, \citenamefont {Haldrup},
  \citenamefont {Glatzel}, \citenamefont {March}, \citenamefont {Doumy},
  \citenamefont {Britz}, \citenamefont {Galler}, \citenamefont {Assefa},
  \citenamefont {Cabaret}, \citenamefont {Juhin}, \citenamefont {van Driel},
  \citenamefont {Kjær}, \citenamefont {Dohn}, \citenamefont {Møller},
  \citenamefont {Lemke}, \citenamefont {Gallo}, \citenamefont {Rovezzi},
  \citenamefont {Németh}, \citenamefont {Rozsályi}, \citenamefont {Rozgonyi},
  \citenamefont {Uhlig}, \citenamefont {Sundström}, \citenamefont {Nielsen},
  \citenamefont {Young}, \citenamefont {Southworth}, \citenamefont {Bressler},\
  and\ \citenamefont {Gawelda}}]{vanko}%
  \BibitemOpen
  \bibfield  {author} {\bibinfo {author} {\bibfnamefont {G.}~\bibnamefont
  {Vankó}}, \bibinfo {author} {\bibfnamefont {A.}~\bibnamefont {Bordage}},
  \bibinfo {author} {\bibfnamefont {M.}~\bibnamefont {Pápai}}, \bibinfo
  {author} {\bibfnamefont {K.}~\bibnamefont {Haldrup}}, \bibinfo {author}
  {\bibfnamefont {P.}~\bibnamefont {Glatzel}}, \bibinfo {author} {\bibfnamefont
  {A.~M.}\ \bibnamefont {March}}, \bibinfo {author} {\bibfnamefont
  {G.}~\bibnamefont {Doumy}}, \bibinfo {author} {\bibfnamefont
  {A.}~\bibnamefont {Britz}}, \bibinfo {author} {\bibfnamefont
  {A.}~\bibnamefont {Galler}}, \bibinfo {author} {\bibfnamefont
  {T.}~\bibnamefont {Assefa}}, \bibinfo {author} {\bibfnamefont
  {D.}~\bibnamefont {Cabaret}}, \bibinfo {author} {\bibfnamefont
  {A.}~\bibnamefont {Juhin}}, \bibinfo {author} {\bibfnamefont {T.~B.}\
  \bibnamefont {van Driel}}, \bibinfo {author} {\bibfnamefont {K.~S.}\
  \bibnamefont {Kjær}}, \bibinfo {author} {\bibfnamefont {A.}~\bibnamefont
  {Dohn}}, \bibinfo {author} {\bibfnamefont {K.~B.}\ \bibnamefont {Møller}},
  \bibinfo {author} {\bibfnamefont {H.~T.}\ \bibnamefont {Lemke}}, \bibinfo
  {author} {\bibfnamefont {E.}~\bibnamefont {Gallo}}, \bibinfo {author}
  {\bibfnamefont {M.}~\bibnamefont {Rovezzi}}, \bibinfo {author} {\bibfnamefont
  {Z.}~\bibnamefont {Németh}}, \bibinfo {author} {\bibfnamefont
  {E.}~\bibnamefont {Rozsályi}}, \bibinfo {author} {\bibfnamefont
  {T.}~\bibnamefont {Rozgonyi}}, \bibinfo {author} {\bibfnamefont
  {J.}~\bibnamefont {Uhlig}}, \bibinfo {author} {\bibfnamefont
  {V.}~\bibnamefont {Sundström}}, \bibinfo {author} {\bibfnamefont {M.~M.}\
  \bibnamefont {Nielsen}}, \bibinfo {author} {\bibfnamefont {L.}~\bibnamefont
  {Young}}, \bibinfo {author} {\bibfnamefont {S.~H.}\ \bibnamefont
  {Southworth}}, \bibinfo {author} {\bibfnamefont {C.}~\bibnamefont
  {Bressler}}, \ and\ \bibinfo {author} {\bibfnamefont {W.}~\bibnamefont
  {Gawelda}},\ }\href {\doibase 10.1021/acs.jpcc.5b00557} {\bibfield  {journal}
  {\bibinfo  {journal} {The Journal of Physical Chemistry C}\ }\textbf
  {\bibinfo {volume} {119}},\ \bibinfo {pages} {5888–5902} (\bibinfo {year}
  {2015})}\BibitemShut {NoStop}%
\bibitem [{\citenamefont {Zhang}\ \emph {et~al.}(2015)\citenamefont {Zhang},
  \citenamefont {Lawson~Daku}, \citenamefont {Zhang}, \citenamefont
  {Suarez-Alcantara}, \citenamefont {Jennings}, \citenamefont {Kurtz},\ and\
  \citenamefont {Canton}}]{JTcanton2015}%
  \BibitemOpen
  \bibfield  {author} {\bibinfo {author} {\bibfnamefont {X.}~\bibnamefont
  {Zhang}}, \bibinfo {author} {\bibfnamefont {M.~L.}\ \bibnamefont
  {Lawson~Daku}}, \bibinfo {author} {\bibfnamefont {J.}~\bibnamefont {Zhang}},
  \bibinfo {author} {\bibfnamefont {K.}~\bibnamefont {Suarez-Alcantara}},
  \bibinfo {author} {\bibfnamefont {G.}~\bibnamefont {Jennings}}, \bibinfo
  {author} {\bibfnamefont {C.~A.}\ \bibnamefont {Kurtz}}, \ and\ \bibinfo
  {author} {\bibfnamefont {S.~E.}\ \bibnamefont {Canton}},\ }\href@noop {}
  {\bibfield  {journal} {\bibinfo  {journal} {The Journal of Physical Chemistry
  C}\ }\textbf {\bibinfo {volume} {119}},\ \bibinfo {pages} {3312} (\bibinfo
  {year} {2015})}\BibitemShut {NoStop}%
\bibitem [{\citenamefont {Dohn}\ \emph {et~al.}(2014)\citenamefont {Dohn},
  \citenamefont {Örn Jónsson}, \citenamefont {Kjær}, \citenamefont {van
  Driel}, \citenamefont {Nielsen}, \citenamefont {Jacobsen}, \citenamefont
  {Henriksen},\ and\ \citenamefont {Møller}}]{dohn2014}%
  \BibitemOpen
  \bibfield  {author} {\bibinfo {author} {\bibfnamefont {A.~O.}\ \bibnamefont
  {Dohn}}, \bibinfo {author} {\bibfnamefont {E.}~\bibnamefont {Örn Jónsson}},
  \bibinfo {author} {\bibfnamefont {K.~S.}\ \bibnamefont {Kjær}}, \bibinfo
  {author} {\bibfnamefont {T.~B.}\ \bibnamefont {van Driel}}, \bibinfo {author}
  {\bibfnamefont {M.~M.}\ \bibnamefont {Nielsen}}, \bibinfo {author}
  {\bibfnamefont {K.~W.}\ \bibnamefont {Jacobsen}}, \bibinfo {author}
  {\bibfnamefont {N.~E.}\ \bibnamefont {Henriksen}}, \ and\ \bibinfo {author}
  {\bibfnamefont {K.~B.}\ \bibnamefont {Møller}},\ }\href {\doibase
  10.1021/jz500850s} {\bibfield  {journal} {\bibinfo  {journal} {The Journal of
  Physical Chemistry Letters}\ }\textbf {\bibinfo {volume} {5}},\ \bibinfo
  {pages} {2414} (\bibinfo {year} {2014})}\BibitemShut {NoStop}%
\end{thebibliography}%

\end{document}